\documentclass[acmsmall]{acmart}
\usepackage[]{hyperref}
\usepackage{xspace}
\usepackage{listings}
\usepackage{cite}
\usepackage{url}

\usepackage{amsmath,amssymb,amsfonts}
\usepackage{amsthm}
\usepackage{algorithmic}
\usepackage{graphicx}
\usepackage{textcomp}
\usepackage{enumitem}
\usepackage{listings}
\usepackage{multicol, multirow}
\usepackage{tikz}
\usepackage{comment}
\usepackage{balance}
\usepackage{pgfplots} 
\usepackage{caption} 
\usepackage{geometry}
\usepackage{xcolor}
\usepackage{relsize}
\usepackage{wrapfig}
\usepackage[]{mdframed}
\usepackage{placeins}

\usepackage{tabularx}

\usepgfplotslibrary{groupplots}

\definecolor{darkgreen}{rgb}{0.0, 0.5, 0.0}

\AtBeginDocument{%
  }

\setcopyright{rightsretained}
\acmDOI{10.1145/3729265}
\acmYear{2025}
\acmJournal{PACMPL}
\acmVolume{9}
\acmNumber{PLDI}
\acmArticle{166}
\acmMonth{6}
\received{2024-11-15}
\received[accepted]{2025-03-06}

\begin{document}

\title{DR.FIX: Automatically Fixing Data Races at Industry Scale}

\author{Farnaz Behrang}
\orcid{0009-0009-6099-7432}
\email{behrang@uber.com}
\author{Zhizhou Zhang}
\orcid{0000-0002-5517-6308}
\email{zzzhang@uber.com}
\affiliation{%
  \institution{Uber Technologies}
  \city{Sunnyvale}
  \country{USA}
}
\author{Georgian-Vlad Saioc}
\orcid{0009-0000-1714-3866}
\affiliation{%
  \institution{Aarhus University}
  \city{Aarhus}
  \country{Denmark}
}
\email{gvsaioc@cs.au.dk}

\author{Peng Liu}
\orcid{0009-0000-7814-9372}
\affiliation{%
  \institution{Uber Technologies}
  \city{New York City}
  \country{USA}
}
\email{peng3141@uber.com}

\author{Milind Chabbi}
\orcid{0000-0003-1021-7644}
\affiliation{%
  \institution{Uber Technologies}
  \city{Sunnyvale}
  \country{USA}
}
\email{milind@uber.com}

\newtheorem{fact}{Fact}

\newcommand{\milind}[1]{{\color{red} Milind: #1}}
\newcommand{\chris}[1]{{\color{orange} Chris: #1}}
\newcommand{\farnaz}[1]{{\color{blue} Farnaz: #1}}
\newcommand{\peng}[1]{{\color{darkgreen} Peng: #1}}
\newcommand{\vlad}[1]{{\color{cyan} Vlad: #1}}
\newcommand{\drfix}{\textsc{Dr.Fix}}

\newcommand{\red}[1]{{\color{red}#1}}
\newcommand{\monorepodevs}{$\sim$6000\xspace}
\newcommand{\noappxmonorepodevs}{6000\xspace}

\newcommand{\numdevs}{thousands\xspace}
\newcommand{\totaldrs}{\red{total data races?}\xspace}
\newcommand{\fixeddrs}{\red{fixed data races}\xspace}
\newcommand{\period}{three\xspace}
\newcommand{\timeperdr}{11 days\xspace}
\newcommand{\monorepoloc}{100 million\xspace}
\newcommand{\monoreposvs}{$\sim$3500\xspace}
\newcommand{\fixeddrrate}{86\%\xspace}
\newcommand{\numdrfixfixed}{193\xspace}
\newcommand{\numdrfixdevs}{a hundred\xspace}
\newcommand{\numcores}{$\sim$2 million\xspace}
\newcommand{\diffsperday}{$\sim$1500\xspace}
\newcommand{\mgo}[1]{\texttt{\small  #1}}

\newcommand{\mgotype}[1]{{\text{\relsize{0}{\tt\color{black}\textbf{#1}}}}}
\newcommand{\mgobuiltin}[1]{{\text{\relsize{0}{\tt\color{black}\textbf{#1}}}}}
\newcommand{\mcode}[1]{{\text{\tt{#1}}}}
\newcommand{\hb}{$HB$}
\newcommand{\drsfound}{$8000$}
\newcommand{\drsfixed}{$4000$}
\newcommand{\dailyraces}{$\textrm{5-15}$}
\newcommand{\rrt}{${\color{red}\blacktriangleright}$}
\newcommand{\grt}{${\color{green}\blacktriangleright}$}
\newcommand{\gplus}{\color{darkgreen}+}
\newcommand{\rminus}{{\color{red}-}}

\newcommand{\testrepeats}{$1000$}
\newcommand{\drfixperiod}{18-month}
\newcommand{\drfixperiodnohyphern}{18 month}

\newcommand{\cureatedbugs}{272}
\newcommand{\cureatedsupersetbugs}{747}

\newcommand{\company}{Uber\xspace}
\newtheorem{Definition}{Definition}[section]
\newtheorem{Theorem}{Theorem}
\newtheorem{Remark}{Remark}
\newtheorem{Observation}{Observation}
\newtheorem{Rule}{Rule}
\newtheorem{Fact}{Fact}
\newtheorem{Lemma}{Lemma}
\newtheorem{Corollary}{Corollary}
\newtheorem{Lcorol}{Corollary}
\newtheorem{Example}{Example}

\lstdefinelanguage{CustomGo}{%
aboveskip=5pt,
belowskip=0pt,
lineskip= {-1.5pt},
language=go,                %
basicstyle=\scriptsize,       %
numbers=left,                   %
numberstyle=\tiny,      %
stepnumber=1,                   %
numbersep=2pt,                  %
backgroundcolor=\color{white},  %
showspaces=false,               %
stringstyle=\scriptsize,
identifierstyle=\scriptsize,
commentstyle=\scriptsize,
basicstyle=\scriptsize\ttfamily,
showstringspaces=false,         %
showtabs=false,                 %
frame=tb,                   %
tabsize=2,                      %
captionpos=b,                   %
breaklines=true,                %
breakatwhitespace=false,        %
title=\lstname,                 %
  sensitive,%
  morecomment=[s]{/*}{*/},%
  morecomment=[l]//,%
  morestring=[b]',%
  morestring=[b]",%
  morestring=[s]{`}{`},%
  morekeywords=[1]{break,case,const,continue,default,defer,%
      else,fallthrough,false,for,func,go,goto,if,import,iota,skip,%
      range,return,select,switch,true,type,nop,%
      var,then,while},%
  morekeywords=[3]{append,cap,close,complex,copy,delete,%
      len,make,new,panic,print,println,recover},%
  morekeywords=[2]{bool,map,byte,complex64,complex128,float32,float64,%
      int,int8,int16,int32,int64,rune,string,interface,struct,%
      uint,uint8,uint16,uint32,uint64,uintptr,chan,error,any},%
  keywordstyle=[1]{\bfseries\color{black}},
  keywordstyle=[2]{\bfseries\color{black}},
  keywordstyle=[3]{\bfseries\color{black}},
  commentstyle=\color{gray},
  backgroundcolor=\color{white},
  escapechar={@},
  moredelim=**[is][\color{darkgreen}+]{+++}{+++},
  moredelim=**[is][\color{red}-]{---}{---},
  numbersep=5pt,
  xleftmargin=10pt,
  numbers=left
}%

\lstdefinelanguage{CustomPyton}{%
aboveskip=5pt,
belowskip=0pt,
lineskip= {-1.5pt},
language=python,                %
basicstyle=\scriptsize,       %
numbers=left,                   %
numberstyle=\tiny,      %
stepnumber=1,                   %
numbersep=2pt,                  %
backgroundcolor=\color{white},  %
showspaces=false,               %
stringstyle=\scriptsize,
identifierstyle=\scriptsize,
commentstyle=\scriptsize,
basicstyle=\scriptsize\ttfamily,
showstringspaces=false,         %
showtabs=false,                 %
frame=tb,                   %
tabsize=2,                      %
captionpos=b,                   %
breaklines=true,                %
breakatwhitespace=false,        %
title=\lstname,                 %
  sensitive,%
  morecomment=[s]{/*}{*/},%
  morecomment=[l]//,%
  morestring=[b]',%
  morestring=[b]",%
  morestring=[s]{`}{`},%
  commentstyle=\color{gray},
  backgroundcolor=\color{white},
  escapechar={@},
  numbersep=5pt,
  xleftmargin=10pt,
  numbers=left,
  morekeywords=[1]{if,for,def,range,retrun,},%
  keywordstyle=[1]{\bfseries\color{black}},
}%

\begin{abstract}
\end{abstract}

\ccsdesc[500]{Software and its engineering~Parallel programming languages}
\ccsdesc[500]{Software and its engineering~Concurrent programming languages}
\ccsdesc[500]{Computing methodologies~Concurrent programming languages}
\ccsdesc[500]{Software and its engineering~Software verification and validation}
\ccsdesc[500]{Information systems~Language models}

\keywords{Data Race, Concurrency, Program Repair, Large Language Models}


\maketitle

Data races are a prevalent class of concurrency bugs in shared-memory parallel programs, posing significant challenges to software reliability and reproducibility.
While there is an extensive body of research on detecting data races and a wealth of practical detection tools across various programming languages, considerably less effort has been directed toward automatically fixing data races at an industrial scale.
In large codebases, data races are continuously introduced and exhibit myriad patterns, making automated fixing particularly challenging.

In this paper, we tackle the problem of automatically fixing data races at an industrial scale. We present \drfix{}, a tool that combines large language models (LLMs) with program analysis to generate fixes for data races in real-world settings, effectively addressing a broad spectrum of racy patterns in complex code contexts. 
Implemented for Go---the programming language widely used in modern microservice architectures where concurrency is pervasive and data races are common—\drfix{} seamlessly integrates into existing development workflows.
We detail the design of \drfix{} and examine how individual design choices influence the quality of the fixes produced. 
Over the past \drfixperiodnohyphern{}s, \drfix{} has been integrated into developer workflows at \company{}
demonstrating its practical utility. 
During this period, \drfix{} produced patches for 224  (55\%) from a corpus of 404 data races spanning various categories; \numdrfixfixed{} of these patches (\fixeddrrate{})  were accepted by more than \numdrfixdevs{} developers via code reviews and integrated into the codebase.

\section{Introduction}
\label{sec:intro}

Data races are a prominent type of concurrency bug in shared-memory parallel programs. 
    They occur when two or more concurrent tasks access the same memory location, and at least one performs a write. 
The programming languages community has extensively studied data races, resulting in a rich body of literature on detecting them via both static~\citep{mayur_pldi_2006, kahlon2007fast, racerd, pratikakis2011locksmith, chatarasi2017extended, engler2003racerx, voung2007relay} and dynamic~\citep{flanagan2009fasttrack, smaragdakis2012sound, predictive_analysis_race, atzeni2016archer, feng1997efficient, cheng1998detecting, savage1997eraser, michael_pldi_2018,yu2005racetrack, marino2009literace, raman2012scalable, sen2008race, serebryany2009threadsanitizer} program analyses. 
Mature tools like ThreadSanitizer~\citep{serebryany2009threadsanitizer} have become industry standards for data race detection in production workflows.

Recent studies have highlighted how concurrency bugs in general~\citep {gobugsTu} and data races in particular~\citep{chabbi2022study} are pervasive in Go programs, due to factors such as the ease of creating concurrency and nuances in programming idioms. 
In a study, Chabbi and Ramanathan~\citep{chabbi2022study} provide a useful classification of a vast corpus of data races based on common causes encountered in an industrial setting. 
Furthermore, they note that new data races are introduced daily alongside the many code changes, at a rate that outpaces developers' ability to address them.

Race detection tools typically stop at reporting the race condition, leaving root cause analysis and remediation to developers. 
\company{} technologies has over \noappxmonorepodevs{} developers working on \monorepoloc{} lines of Go code. 
We observe \dailyraces{}~new data races per day, across \diffsperday{} revisions, using the Go's built-in ThreadSanitizer~\citep{serebryany2009threadsanitizer} dynamic race detector.
Our experience with \numdevs{} of developers over \period{} years at \company{} revealed that typical developers: 
(a) are not concurrency experts; 
(b) need assistance to address data races; and 
(c) spend an average of \timeperdr{} of wall-clock time
fixing a single data race. 
This highlights a substantial loss in productivity due to complex concurrency bugs like data races, which are not the main focus of developers working on business logic.

These observations naturally call for an automated approach to fixing data races. 
However, automatically resolving any software defect is challenging, especially concurrency bugs. 
Prior efforts to fix data races have been limited to structured parallel programs~\citep{surendran2014test}, often introducing heavy-weight synchronization such as locks or yields around racing statements~\citep{hippoCostea, surendran2014test, jin2012automated, KhoshnoodConcBugAssist, HealingKrena,lin2018pfix}, working on small code scopes~\citep{liu2016understanding}, and lacking evaluation at an industrial scale~\citep{hippoCostea, surendran2014test, lin2018pfix}.

A major challenge in rule-based data-race remediation is that these races seldom conform to a single algorithmic formula; while many fall into general patterns, each instance may demand its own tailored solution \citep{chabbi2022study, liu2016understanding, GuangpuLargeStudy}.
Naively using locks around racy accesses can introduce performance losses~\citep{tallentLockContention, toddlerNistor, GuanchengCriticallockanalysis, boyd2012non}, or result in incorrect fixes, which may be harder to detect~\citep{LuAVIO, lu2008learning, park2009ctrigger}. 
An effective fix should retain the original intent of the programmer while eliminating the race.
Solutions may involve making a copy of an object, moving code to avoid racy accesses, correctly placing synchronization constructs, or using appropriate data structures (e.g., concurrent maps) or atomic variables, to name just a few. 
The necessary analysis and transformation can be intra-procedural or inter-procedural; within a single file or across multiple files and packages. 
A robust, industry-scale race-fixing system must address all these scenarios and provide highly accurate fixes within typical developer workflows, adhering to the language and codebase's coding style and idioms.

In this work, we leverage generative AI (GenAI) to bridge the gap between detected and unresolved data races in industrial applications.
While GenAI excels in understanding code within large contexts and generating content, it alone cannot deliver the precision required for complex concurrency bug fixes. 
To harness its strengths, we developed a hybrid approach that integrates decades of program analysis research with GenAI’s capabilities in reasoning across intricate codebases.

A naive application of large language models (LLMs) directly to code often results in simplistic fixes---such as blanket locks---that overlook performance and intent, compromising both accuracy and efficiency. 
Our system combines retrieval-augmented generation (RAG)~\citep{lewis2021retrievalaugmentedgenerationknowledgeintensivenlp} with a concurrency-focused code skeleton abstraction to ensure the LLM applies past fixes in a structurally accurate context. 
This RAG-approach guides the LLM to resolve new bugs by drawing on previously addressed, similar issues, adapting them to the current context with precision.
Given that industrial codebases are dense with domain-specific logic and terminology, standard retrieval methods tend to prioritize business logic over concurrency patterns. 
Our system counters this by using a novel code abstraction technique that reduces noise, emphasizing concurrency structure over business logic. 
Furthermore, our iterative retrieval approach examines solutions at multiple levels of program abstraction, enabling it to address a wide range of scenarios effectively.

The resulting system---\drfix{}---effectively addresses real-world data races at an industrial scale. 
The input to \drfix{} is the code repository and a data race report generated by a detection tool like Go's built-in ThreadSanitizer. 
The output is a code patch fixing the data race, which is then sent to developers for review. 
\drfix{} can fix data races across various categories, demonstrating versatility in both inter- and intra-procedural contexts, and across files and packages. 
The proposed fixes are not limited to merely protecting racy accesses with locks; instead, they provide idiomatic solutions encompassing all the aforementioned strategies and more.

Over the \drfixperiod{} observation period, usage of \drfix{} steadily increased as its capabilities evolved. In total, \drfix{} successfully produced fixes for 224 (55\%) of 404 identified data races. Among these 224 patches, \numdrfixfixed{} were approved following reviews by over \numdrfixdevs{} developers, yielding a noteworthy acceptance rate of \fixeddrrate{}. Moreover, tickets resolved by \drfix{} were completed in an average of three days, compared to 11 days for non-\drfix{} tickets. Finally, a user survey indicated a 67.6\% positive sentiment among those who reviewed \drfix{}-generated revisions.

In this paper, we make the following contributions:
\begin{itemize}
    \item Develop a first-of-a-kind, automatic, data race fixing system---\drfix{}---for Go. 
    \drfix{} combines program analysis with Generative AI to deliver targeted and idiomatic fixes. \drfix{} develops novel abstraction techniques to denoise example retrieval, allowing it to adaptively address complex concurrency issues in large-scale Go codebases. 
    \item Demonstrate the practical applicability and usefulness of \drfix{} by deploying it in \company{}'s Go codebase, delivering automated fixes for \numdrfixfixed{} data race with an acceptance rate of \fixeddrrate{}, and 
    \item Empirically evaluate \drfix{} through systematic ablation studies to highlight the necessity and importance of each component in its design.
\end{itemize}

Although our deployment and evaluation of \drfix{} are focused on the Go programming language within \company{}, the underlying ideas and principles apply to other languages and systems. 
To demonstrate this generality, we also applied \drfix{} to automatically fix data races in Swift-based iOS mobile applications on a smaller scale, achieving similar success.

The remainder of the paper is structured as follows. Section~\ref{sec:bg} outlines the necessary background. Section~\ref{sec:method} introduces the high-level methodology behind \drfix{}. Section~\ref{sec:impl} delves into the implementation details. Section~\ref{sec:eval} presents a detailed evaluation of the approach. Related work is discussed in Section~\ref{sec:rel}, and Section~\ref{sec:concl} concludes the paper.

\section{Background}
\label{sec:bg}

In this section, we briefly introduce the concurrency in Go,  characterize the nature of the Go code at \company{}, and describe the data race detection system at \company{}.

\subsection{Concurrency in  Go}
Go is a statically typed, garbage-collected, and compiled language with first-class support for concurrency, making it ideal for scalable network services and distributed systems. 
Goroutines in Go are lightweight, runtime-managed threads, which share the same address space and can communicate through both shared memory and message passing (via channels). 
Goroutines are created by prefixing a function call with the \mgo{go} keyword.
Go does not enforce structured parallelism~\citep{openmp, charles2005x10, HJcave, cilkFrigo}; for example, a child goroutine can outlive its parent.
Go's memory model~\citep{gommem:online} is based on the ``happens-before'' relationship, but it is less explicit than C++~\citep{boehm2008foundations} or Java~\citep{manson2005java}.
A happens-before (\hb{}) relationship exists between the parent goroutine, creating a child goroutine.
Go provides multiple synchronization mechanisms, some of which we discuss below:

\begin{itemize}
    \item \textbf{Channels:} 
    \mgo{ch <- item} is the syntax to send an \mgo{item} on a channel \mgo{ch}. 
    Receiving an item from a channel \mgo{ch} is represented as \mgo{<-ch}.
    Channels enforce an \hb{} relationship between the sender and the receiver goroutines.    
    \item \textbf{sync.WaitGroup:} The \mgo{WaitGroup} is used to wait for a collection of goroutines to complete. It provides the methods \mgo{Add(int)}, \mgo{Done}, and \mgo{Wait}.
\mgo{Add} increments the counter by an integer, typically the number of goroutines to wait for.
\mgo{Done} decrements the counter, signaling the completion of a goroutine.
\mgo{Wait} blocks until the counter reaches zero, ensuring that all goroutines have completed, which enforces an \hb{}  relationship between pairs of goroutines performing a \mgo{Done} with the matching goroutine performing the \mgo{Wait}.
    \item \textbf{sync.Atomic}: Go lacks built-in atomic/volatile variables. One can apply atomic operations on certain primitive types. All atomic operations behave in a sequentially consistent order.

\item \textbf{sync.Mutex:} Mutex objects introduce \mgo{Lock/Unlock} APIs. There is an \hb{}  relationship between the goroutine performing an \mgo{Unlock} with the goroutine performing the subsequent \mgo{Lock} on the same mutex.   

\end{itemize}

Sharing memory between goroutines is a very accessible Go feature. 
Commonly, this is achieved by using
\textbf{capture-by-reference in lambdas}. Go closures (anonymous functions) capture all free variables by reference, which can lead to unexpected behaviors in concurrent programs.
Another Go feature with subtle concurrent behavior consists of
\textbf{built-in maps.} The built-in type \mgo{map} can be defined with a given key and value type. Maps in Go are not thread-safe.
All other shared heap objects are subject to concurrency bugs, as in other languages.
For additional details refer to~\citep{chabbi2022study}.

\subsection{Characteristics of \company{}'s Go Code}
\begin{wraptable}{r}{0.4\textwidth}
    \caption{Salient aspects of the Go codebase at \company{}.}
    \label{tab:monorepofeatures}
  \centering
\scriptsize
    \begin{tabular}{r|c|c|c}
           & \textbf{Total} & \textbf{Product} & \textbf{Test} \\
        \hline
        \hline
        \multicolumn{4}{c}{\textbf{Entire \company{} monorepo} (159K packages)}
        \\
        \hline
        \textbf{Files} & 382K & 245K & 137K\\
        \textbf{Lines of code} & 97.2M & 59.3M & 37.9M
        \\
        \hline
        \multicolumn{4}{c}{\textbf{Including concurrency features} (24K packages)}
        \\
        \hline
        \textbf{Files} & 53K & 28K & 25K
        \\
        \textbf{Lines of code} & 15.6M & 6.2M & 9.4M
    \end{tabular}

\end{wraptable}

\company{} is a software service company that has adopted Go as its primary language for developing both its user-exposed real-time query processing web services (microservices) and internal tools. 
The company has \monorepodevs{} active Go developers with varying degrees of expertise.
The Go codebase is hosted in a single repository, which has about \monorepoloc{} lines of code,
and receives \diffsperday{} daily revisions.
The \monoreposvs{} Go microservices spanning the codebase are deployed across many hosts.
Table~\ref{tab:monorepofeatures} highlights some salient features of the codebase, which includes only internally developed code and excludes any external  dependencies.
At runtime, Go services at \company{} typically feature $\sim 2000$ goroutines~\citep{chabbi2022study, saioc2025dynamic, saioc2024unveiling}.

\subsection{Data Races in Go}
Data races in Go occur between two goroutines for the same fundamental reasons as in other languages. 
Data races can lead to inconsistent views of shared data as updates may not be visible across threads; the interleaving of operations can create unexpected dependencies on the execution order.
Data races often result in hard-to-reproduce bugs and undefined program behavior, where the program's outcome becomes unpredictable—producing inconsistent results, unexpected crashes, and corrupted data, sometimes introducing security vulnerabilities~\citep{mckenney2010memory,impervaVulnerability:online}. 
At \company{}, we have observed data races leading to array-out-of-bounds errors and runtime panics, among others.

We use Listing~\ref{lst:motivation} to demonstrate a simple, yet frequently occurring data race in Go. 
On line~\ref{motovation:init}, the \mgo{:=} operator declares and assigns a value to variable \mgo{err}. 
Line~\ref{motovation:wg} creates a \mgo{WaitGroup} that is incremented once on line~\ref{motovation:wgadd}. 
The parent goroutine then creates a child goroutine on line~\ref{motovation:go1}, followed by an error check on line~\ref{motovation:race2}, reusing the \mgo{err} variable.
If no error is returned by \mgo{Task2}, the parent waits for the termination of the child at line~\ref{motovation:wait}.
Meanwhile, the child prepares a call to \mgo{Done} at the end of its execution by using the \mgo{defer} keyword, followed by a local error check at line~\ref{motovation:race1}.

The issue lies at line~\ref{motovation:race1}, where the developer mistakenly assigns to the \mgo{err} variable with an \mgo{=}, capturing \mgo{err} by reference from the enclosing scope. 
The sharing causes a write-write data race between the two goroutines on lines~\ref{motovation:race1} and ~\ref{motovation:race2}.

The data race can be solved, without any synchronization, by simply locally re-declaring \mgo{err} within the child with \mgo{:=} instead of \mgo{=}, as shown in Listing~\ref{lst:motfix} on Line~\ref{motfix:1}.
This makes \mgo{err} a fresh, lexically-scoped variable, instead of a name captured by reference from the parent scope.

\subsection{Data Race Detection in Go and Adaptation at \company{}}
Dynamic analysis~\citep{biswas2015valor,feng1997efficient,flanagan2009fasttrack,mathur2018happens,pozniansky2003efficient,raman2012scalable,samak2015synthesizing,smaragdakis2012sound}
is a popular technique to identify data races. 
Go provides a built-in race detector that instruments the code at compile time and detects races during execution via the ThreadSanitizer~\citep{serebryany2009threadsanitizer} runtime library. 
The race report (Figure~\ref{fig:cct}) contains the stack trace of each unordered access, alongside the source function name, file, and line number, as well as the stack trace of the parent goroutine (if any) at the creation point of the racy goroutine.

At \company{, we exercise our test suites with data race detection enabled daily, filing tickets for newly found bugs similar to the system described in~\citep{chabbi2022study}. 
Over the past three years, this system has identified over \drsfound{} data race defects.
The system continues to find \dailyraces{} new data races\footnote{Some data races may share the same underlying cause, but exhibit different execution traces}  every day, either in newly written code or by exercising alternative execution schedules\footnote{Recently, we have started blocking changes containing an unseen data race as part of the continuous integration system, despite the potential inconvenience where an otherwise benign change may exercise a preexisting race.}.
A team of experts supports developers by helping them understand data race reports, offering guidance on fixing races, and even providing expert-level code fixes.
However, the rate of fixing lags behind the rate of discovery, necessitating automation.

The \drfix{} automation described in Section~\ref{sec:method} easily fixes the example shown in Listing~\ref{lst:motivation}.
While this specific example is small and could be analyzed and fixed with simple program analysis, more complex races exist that our tool, \drfix{}, can address. We cover several such cases in Section~\ref{sec:eval}.

\begin{figure}[!t]
  \centering
  \begin{minipage}[t]{0.48\linewidth}
    \begin{lstlisting}[language=CustomGo, label={lst:motivation},
    caption=Example of a write-write data race caused by assignments to a variable captured by reference.,
    captionpos=b]
 func SomeFunction(...) error {
@\label{motovation:init}@   err := someWork()
   if err != nil {
     return err
   }
@\label{motovation:wg}@   var wg sync.WaitGroup
@\label{motovation:wgadd}@   wg.Add(1)
@\label{motovation:go1}@   go func() {
@\label{motovation:done2}@     defer wg.Done()
     // 'err' is captured by reference
@\label{motovation:race1}\rrt@    if err = Task1(...); err != nil {
       ...
     }
   }()
@\label{motovation:race2}\rrt@ if err = Task2(...); err != nil { ... }
@\label{motovation:wait}@   wg.Wait()
   ...
 }
\end{lstlisting}
  \end{minipage}
  \hfill
  \begin{minipage}[t]{0.49\linewidth}
    \begin{lstlisting}[language=CustomGo, label={lst:motfix},
    caption=The data race is fixed via variable redeclarations.,
    captionpos=b]
 func SomeFunction(...) error {
   err := someWork()
   if err != nil {
     return err
   }
   var wg sync.WaitGroup
   wg.Add(1)
   go func() {
     defer wg.Done()
     // 'err' is a fresh declaration
@\label{motfix:1}\grt@    if err := Task1(...); err != nil {
       ...
     }
   }()
@\label{motfix:2}\grt@ if err = Task2(...); err != nil { ... }
   wg.Wait()
   ...
 }
    \end{lstlisting}
  \end{minipage}
\end{figure}

\section{\drfix{} Methodology}
\label{sec:method}

Generative AI with large language models (LLMs) ~\citep{radford2018improving, radford2019language, brown2020language, vaswani2023attentionneed, kasneci2023chatgpt, achiam2023gpt, yao2024tree, wei2022emergent} opens new avenues for addressing complex challenges such as fixing data races. 
LLMs can analyze code snippets, identify similar data races by mapping them to close vectors in high-dimensional space, and propose candidate solutions based on analogous, previously resolved issues. 
This approach shifts the focus from manually developing repair algorithms~\citep{jin2012automated, KhoshnoodConcBugAssist, HealingKrena,lin2018pfix, hippoCostea, surendran2014test} to appropriately prompting LLMs with relevant information to fix a data race; appropriate prompting in turn relies on 
curating and selecting high-quality examples of data race fixes that guide the model in generating new solutions. 
Such an approach enables broader coverage in resolving diverse patterns of data races, where solutions may involve inspecting and modifying extensive code contexts.
In industrial environments, maintaining a rich set of past data race resolutions and thoroughly validating changes through testing and code reviews is both feasible and practical. 
Building on these principles, our methodology involves the following key steps:

\subsection{Step 0: Database Creation with Embedding}
To support the retrieval of past solutions, we maintain a database \( \mathcal{D} \) containing previously encountered data races.
Each entry in \( \mathcal{D} \) includes the code item representing the racy code and its fix, which is then bound to the skeletonized form of the buggy code item embedded into a high-dimensional vector space, for similarity-based retrieval.
We aggregate all code items found across every data race to construct the database as follows:
\[
\mathcal{D} = \{ E(S(b_k)) \mapsto (b_k, f_k) \mid k = 1, 2, \dots, m\}
\]
where:
\begin{itemize}
    \item \( b_k \in \mathcal{B} \) is the buggy code item, where \( \mathcal{B} \) is the domain of all possible buggy code items,
    \item \( f_k \in \mathcal{F} \) is the corresponding fixed code item, where \( \mathcal{F} \) represents the domain of all valid fixes, and
    \item \( E(S(b_k)) \) is the \( d \)-dimensional embedding of the skeleton \( S(b_k) \), capturing the essential concurrency structure and serving as the vector for similarity-based retrieval.
\end{itemize}

Here, \( S(b_k) \) represents the simplified ``skeleton'' of \( b_k \), created by retaining only the concurrency constructs and the race-relevant variables while omitting unrelated details. This skeleton is then mapped to a vector \( E(S(b_k)) \) in high-dimensional space, which allows us to efficiently retrieve relevant entries based on structural similarity.

\subsection{Step 1: Contextual Information Extraction}
For each new data race report \( r \), we extract contextual code elements essential to understand the race. These elements are organized as an ordered set:
\[
    I(r) = (i_1, i_2, \dots, i_n)
\]
Each item, \( i_j \), in $I(r)$ represent aspects of \( r \), such as: (a) the source code of the functions directly involved in the race, (b) relevant portions of files, (c) thread creation points, or (d) the lowest common ancestor function of the racing threads.
Each item \( i_j \) contributes to a broader context, and aids in accurate fix placement.

\subsection{Step 2: Skeleton Creation and Embedding for New Race Reports}
To efficiently retrieve relevant examples from the database, each code item \( i_j \in I(r) \) for the new data race report undergoes skeleton creation and embedding, similar to database creation.

\begin{enumerate}
    \item \textbf{Skeleton Creation:} We generate a skeleton \( S(i_j) \) for each code item \( i_j \), isolating the concurrency constructs and race-related variables while omitting irrelevant details. This creates a concise representation focused on the concurrency structure, which mirrors the skeletonization process used for items in the database. 
    See Section~\ref{sec:skeleton} for the implementation.
    
    \item \textbf{Embedding:} Each skeleton \( S(i_j) \) is then transformed into a \( d \)-dimensional vector \( e_{i_j} = E(S(i_j)) \in \mathbb{R}^d \). This vector serves as the search key, allowing us to find similar entries in \( \mathcal{D} \) based on structural similarity.
\end{enumerate}

\subsection{Step 3: Matching and Fix Generation}
\label{sec:match-and-fix}
Once embedded, we utilize retrieval-augmented generation (RAG)~\citep{lewis2021retrievalaugmentedgenerationknowledgeintensivenlp} to match and generate a fix for each \( i_j \) in a predetermined order:

\begin{enumerate}
    \item \textbf{Best Match Retrieval:} For each vectorized skeleton \( e_{i_j} \), we retrieve the closest match from the database using a similarity function \( \phi : \mathbb{R}^d \times \mathbb{R}^d \to \mathbb{R} \):
    \[
    e^*_{i_j} = \arg \max_{e_k \in D} \phi(e_{i_j}, e_k)
    \]
    This yields the corresponding database entry \( (b^*_{i_j}, f^*_{i_j}, E(S(b^*_{i_j}))) \), from which we retrieve both the original buggy code \( b^*_{i_j} \in \mathcal{B} \) and its fix \( f^*_{i_j} \in \mathcal{F} \).
    
    \item \textbf{Fix Generation:} Using the transformation model \( M \), we generate a candidate fix \( f_{\text{cand}} \) by combining \( i_j \), the retrieved buggy item \( b^*_{i_j} \), and its fix \( f^*_{i_j} \):
    \[
    f_{\text{cand}} = M(i_j, b^*_{i_j}, f^*_{i_j}, h)
    \]
    where \( h \in \mathcal{H} \) represents optional feedback from prior attempts. Although \( f_{\text{cand}} \) is derived from prior examples, it represents a new fix within the domain \( \mathcal{F} \), allowing for novel solutions that extend beyond previous fixes. Furthermore, $b^*_{i_j}$, $f^*_{i_j}$, and $h$ can be \texttt{null}.
     
    \item \textbf{Validation and Iteration:} We validate \( f_{\text{cand}} \) by using a validation function $V$, defined as:
    \[ 
        V : \mathcal{F} \to \{ \text{True, False} \} \times \mathcal{H},\textrm{ where } V(f_{cand}) = (b, h')
    \]
    We have that $b = True$ whenever \( f_{\text{cand}} \) resolves the data race. If validation fails, the attempt feedback \( h' \) informs the next iteration at code item $i_{j+1}$.
    The process repeats until the first successful validation, or until all code items have been exhausted.
\end{enumerate}

This methodology leverages the prior knowledge captured in \( \mathcal{D} \) while also enabling the generation of new fixes through the LLM-based model \( M \). 
Moreover, the model does not merely replicate past fixes but uses them as references to generate appropriate solutions, potentially creating novel fixes that have not been previously encountered. 
The embedding of skeletons into a vector space especially allows finding relevant fixes in case of superficial code differences.
The validation of generated fixes before acceptance ensures that they indeed resolve the reported data race.

\section{\drfix{} Implementation}
\label{sec:impl}

\lstset { %
basicstyle=\ttfamily\footnotesize,
    keywordstyle=\normalfont,
    morekeywords={},
    deletekeywords={int, return, if, else, for, while, map,go}
}

Figure~\ref{fig:overview} depicts the implementation of the race-fixing system deployed at \company{}, including key components and their internals.
Table~\ref{tab:mapping} maps the concepts presented in Section~\ref{sec:method} to the implementation components.
Each substantially impacts the quality of results, shown via evaluations in Section~\ref{sec:eval}.

\begin{figure}[!t]
  \centering
  \includegraphics[width=.8\linewidth]{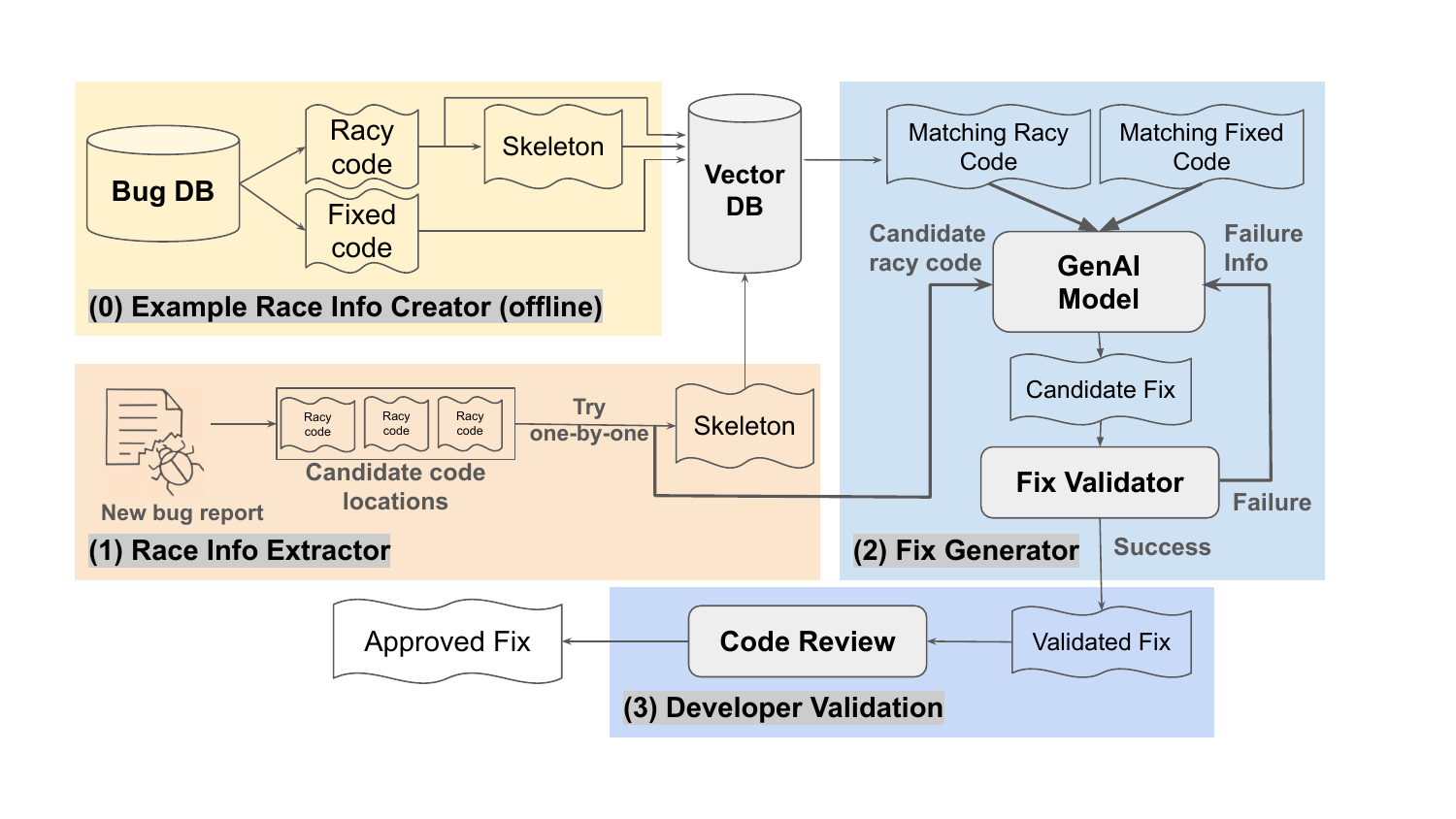}
  \caption{\drfix{} schematic diagram.}
  \label{fig:overview}
\end{figure}
\begin{table}[!t]
\scriptsize
    \caption{Description of individual components used in \drfix{}.}
    \label{tab:mapping}

    \centering
    \begin{tabular}{r|l|l}
         Component &  Choice & Description \\ \hline
          Data store $\mathcal{D}$& ChromaDB~\citep{Chroma2:online} & Vector database \\
          Skeletonization $S$ & AST-based program slicing~\citep{weiser1984program, xu2005brief, binkley1996program} & Preserves concurrency constructs and racy variables. \\
          Embedding $E$ & Sentence Transformers~\citep{reimers-2019-sentence-bert} \texttt{all-MiniLM-L6-v2} & Converts skeletons to $d$-dimensional vector\\
          Similarity $\phi$ & Cosine similarity & Retrieves examples close in semantics to a query vector. \\
          Model $M$ & ChatGPT 4.0 Turbo~\citep{Newmodel93:online}, 4o~\citep{HelloGPT26:online}, and o1-preview~\citep{Introduc16:online} & State-of-the-art LLM models with 128K token capacity.\\
          Extra params $\mathcal{H}$ & Past context and failure info. & Chain of thought~\citep{yao2024tree, wei2022chain} to guide the next step.\\
         Validator $V$ &  package-level tests run \testrepeats{} times & Checks fixes for correctness.\\         
    \end{tabular}
\end{table}

\subsection{Example Race Info Creation}\label{sec:example}
At \company{}, we maintain a database of high-quality data race fixes.
Currently, it has \cureatedbugs{} examples selected from a total of \drsfixed{} fixed data races.
The selection process picks past data races fixed by selected Go concurrency experts over a two-year git revision history.
Frequently encountered racy patterns~\citep{chabbi2022study} appear more often in the dataset, as shown in Table~\ref{tab:category} in Section~\ref{sec:eval}.

\drfix{} extracts the raw source code \emph{before} and \emph{after} the fix, which it then translates to a skeleton (Section~\ref{sec:skeleton}). 
The extracted skeleton code is converted into a $d$-dimensional vector representation using the sentence transformer embedding model \texttt{all-MiniLM-L6-v2}\citep{reimers-2019-sentence-bert}. These vectors are then stored in a vector database (ChromaDB\citep{Chroma2:online}), where each key is a vector and the associated value is a tuple containing the racy code and its corresponding fix. 
When querying arbitrary vectors, we use cosine similarity to find the nearest match among the stored vectors and retrieve relevant examples.
Populating the database is a one-time activity that is typically very fast (several minutes). 
We update the database periodically as new items are found.

\subsection{Race Info Extraction}

Figure~\ref{fig:cct} illustrates an example of calling contexts that access the same data captured for a data race bug. 
The report also includes the stack trace of the parent goroutine at the point where the child was spawned. 
Each stack frame contains information about the function, file, and line number, which we use to extract the source code from the codebase. 

\begin{figure}[!t]
    \centering
    \includegraphics[width=0.8\linewidth]{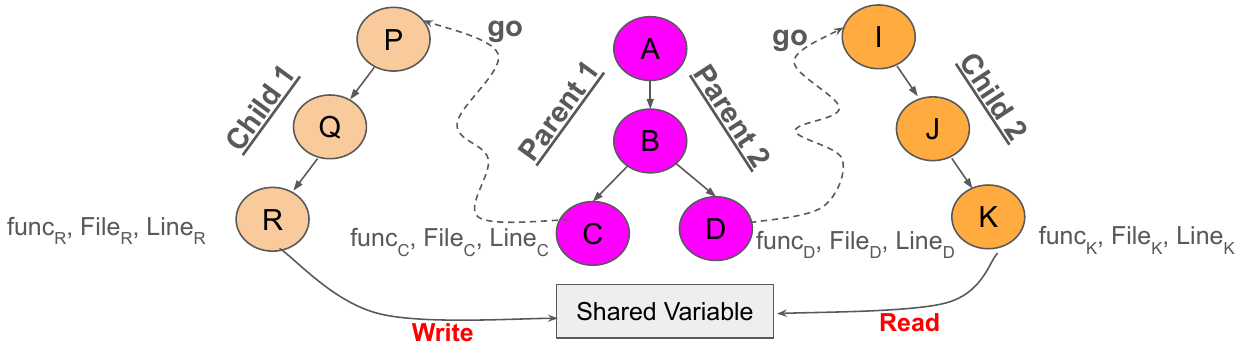}
    \caption{Contextual information of a race report with two racing goroutines, Child1 and Child2. The call paths of Child1 and Child2 at the time of the race are $P\rightarrow Q \rightarrow R$ and $I\rightarrow J \rightarrow K$, respectively. Child1 was created by Parent1 at the call path $A\rightarrow B \rightarrow C$; Child2 was created by Parent2 at call path $A\rightarrow B \rightarrow D$. Each node in the call graph carries its function name and source location. Racy accesses have read/write annotations.}
    \label{fig:cct}
\end{figure}
\paragraph{Fix location:} The \textit{Race Info Extractor} analyzes these calling contexts to extract the following potential fixing locations in the source code:

\begin{enumerate}
    \item The \textbf{leaf functions} (\mgo{raceInfo.leaf}) in the stack traces of the two goroutines involved in the data race.
    This is often the obvious fix candidate, requiring only limited context.
    \item The \textbf{test function} (\mgo{raceInfo.test}) that exercised the race.
    This is typically the root frame of one of the parent stacks. 
    This choice is driven by our observation that concurrently running independent tests that share the same objects can lead to data races not expected by the code under test. 
    In these situations, fixing the test may be preferable to making the tested code thread-safe, unless it is specifically intended as a thread-safe library.
    \item The \textbf{Lowest Common Ancestor (LCA) function}  (\mgo{raceInfo.lca}) in the ancestry of the two goroutines involved in the race. 
    The LCA represents the common point of concurrency creation, before which execution was serial. 
    Hence, it offers opportunities~\citep{surendran2014test,feng1997efficient,raman2012efficient} to control concurrency, make copies, or create and share necessary synchronization objects needed by both goroutines. 
    In Figure~\ref{fig:cct}, the node $B$ is the LCA of the two racing goroutines.
\end{enumerate}

\paragraph{Fix scope:} We maintain two source code scopes for each of these code locations---the contents of the function(s) \mgo{raceInfo.func} and the contents of the entire file(s) \mgo{raceInfo.file}.
We include the entire file to handle cases where the fix needs to change multiple places in the code, such as other functions or data structures declared elsewhere in the files.
We limit our scope to only two files, which need not be in the same package.
The function names from a bug stack trace form a stable hash, which is later used to check if a fix has eliminated the data race.

\subsection{Code Skeleton Creation}
\label{sec:skeleton}
Given a Go source file and a specific line number(s) where data race occurs, this step in \drfix{} performs a specialized program slicing~\citep{weiser1984program} focused on concurrency constructs and shared variables.
It is necessary to perform this step since code at industry scale is standardized. It follows consistent style and naming conventions. Simply matching the raw source code can lead to inaccurate result since the common library and same code repository prefix can add noise and very likely they are not adding value to data race fixing context. 
In order to fix data races, we want to filter out as much irrelevant information as possible and only retain relevant components.
 
The approach starts by parsing the Go source files into its Abstract Syntax Tree (AST).
It then identifies the function(s) containing the data race. 
It uses the variable names found on the lines involved in race as the shared \emph{variables of interest}.
It identifies statements containing constructs such as  \mgo{go}, \mgo{WaitGroup}, \mgo{sync}, \mgo{Lock}, \mgo{Unlock},  \mgo{atomic}, and \mgo{channels} as relevant \emph{concurrency constructs.}
 
It then simplifies the AST by recursively traversing function bodies, keeping control structures like loops and conditionals only if they transitively contain relevant concurrency constructs or variables of interest. 
The variables of interest are consistently renamed (e.g., prefixed with \mgo{racyVar}) to highlight their significance. 
The remaining non-essential variables are renamed with simple names like \texttt{v1}, \texttt{v2}.
The outcome is a distilled version of the original functions that highlights the core concurrency interactions leading to the data race. 

\begin{figure}[!t]
  \centering
  \begin{minipage}[b]{0.48\linewidth}
    \begin{lstlisting}[language=CustomGo, label={lst:beforeSkeleton}, caption=A Go code involved in a data race on variable \mgo{err}., captionpos=b]
@\label{noskeletonEntry}@func (s *storeObject) ProcessStoreData(
  ctx *Context, req *Request) error {
@\label{noskeletonFiestcall}@  err := s.Validate(request)
@\label{noskeletonif}@  if err != nil {
     return err
@\label{noskeletonifend}@  }
  var bazaarStores BazaarStores
  var uuidDefectRateMap UUIDMap
@\label{noskeletonGo1}@  group.Go(func() error {
    docs := s.GetNecessaryDocs()
@\label{noskeletonifdropped}@    if flipr.GetBool(xpAdditionalDocs) {
      otherDocs := s.GetAdditionalDocs()
      docs = append(docs, otherDocs)
@\label{noskeletonifdroppedend}@    }
@\label{noskeletonrace1}@    bazaarStores, err = 
      s.LoadStores(ctx, req, docs)
    return err
    })
@\label{noskeletonGo2}@  group.Go(func() error {
@\label{noskeletonrace2}@    uuidDefectRateMap, err = 
      s.LoadOAData(ctx, s.DocstoreClient, req)
    return err
  })
@\label{noskeletonWait}@  err = group.Wait()
  ...
}
    \end{lstlisting}
  \end{minipage}
  \hfill
  \begin{minipage}[b]{0.48\linewidth}
    \begin{lstlisting}[language=CustomGo, label={lst:afterSkeleton}, caption=Listing~\ref{lst:beforeSkeleton} transformed to its concurrency skeleton using program slicing., captionpos=b]
@\label{skeletoonEntry}@func (v1 *type1) func1(
  v2 *type2, v3 *type3) type4 {
@\label{skeletonFiestcall}@  racyVar1 := v1.func2(v3)
@\label{skeletonif}@  if racyVar1 != nil {
      return racyVar1
@\label{skeletonifend}@  }
  var v5 type5
  var v6 type6
  v7.Go(func() type4 {





    v5, racyVar1 = 
      v1.func3(...)
    return racyVar1
  })
  v7.Go(func() type4 {
    v6, racyVar1 = 
      v1.func4(...)

    
    return racyVar1
  })
  racyVar1 = v4.Wait()
  ...
    \end{lstlisting}
  \end{minipage}
\end{figure}

Listing~\ref{lst:beforeSkeleton} shows an example program that contains a data race on the \mgo{err} variable due to concurrent writes on lines~\ref{noskeletonrace1} and~\ref{noskeletonrace2}.
Its corresponding skeleton is shown in Listing~\ref{lst:afterSkeleton}.
Notably, the variable of interest \mgo{err} is renamed to \mgo{racyVar1}.
The if-then control structure on lines~\ref{noskeletonif}-\ref{noskeletonifend}, which accesses \mgo{err} in Listing~\ref{lst:beforeSkeleton}, is retained in Listing~\ref{lst:afterSkeleton}.
The two goroutine launch constructs and the wait statements, respectively, on lines~\ref{noskeletonGo1}, \ref{noskeletonGo2}, and~\ref{noskeletonWait} from Listing~\ref{lst:beforeSkeleton} are retained in Listing~\ref{lst:afterSkeleton}.
However, the if-then block in the first closure (lines~\ref{noskeletonifdropped}-\ref{noskeletonifdroppedend} in Listing~\ref{lst:beforeSkeleton}), is eliminated in Listing~\ref{lst:afterSkeleton} because it does not have any concurrency constructs and the block does not touch any variables of interest.

\subsection{Fix Generation}
Fix generation is an iterative process that aims to address potential code locations identified during the \emph{Race Info Extraction} step. 
Listing~\ref{lst:fixgeneration} in Appendix~\ref{sec:pseudocode}, submitted as additional material with this paper, illustrates the procedure to produce a fix. 
Specifically, we attempt the following six \emph{fix-location} and \emph{fix-scope} combinations: \texttt{[raceInfo.test, raceInfo.leaf, raceInfo.lca]} crossed with \texttt{[raceInfo.func, raceInfo.file]}.

Modern large language models (LLMs) have demonstrated high-quality content generation when provided with relevant examples, a technique often referred to as few-shot learning~\citep{brown2020language}. 
For a given race information (location, scope) tuple, \drfix{} first retrieves its structural skeleton. 
It then fetches the most relevant previously found bug and its corresponding fix from the curated vector database. 
Using this information, it dynamically constructs a prompt.
Crucially, we always include a special case of \emph{empty example}, among the set of examples we attempt.
The empty example enables the LLM to attempt a fix from its inherent capabilities without being guided by a specific example.
The empty example is valuable when the RAG lacks a relevant fix in its database, but the LLM might have a better example in its training set.
In our evaluation, we observe that the modern LLMs can fix 47\% of data races with their inherent capabilities by being trained on the world-wide-web of information.
However, we also see that our curated examples elevate the fix rate to 66\%. 

Our choice to fix the defect is the following LLMs---ChatGPT 4.0 Turbo~\citep{Newmodel93:online}, 4o~\citep{HelloGPT26:online}, and o1-preview~\citep{Introduc16:online}; however, the model can be easily substituted for alternatives. 
Appendix~\ref{sec:prompt} in the supplemental material shows an example prompt we craft for a data race.

The model response is parsed into an AST; depending on the fix-scope, we either insert the fix into the existing code via AST rewrites, or replace the files in bulk.

\subsubsection{Fix Validation}
After patching the code with LLM-generated fix, \drfix{} attempts to build and run each test from the patched packages 1,000 times. 
Validation succeeds if the tests consistently demonstrate eliminating the data race in all runs. 
Any failures during compilation or testing, or if the data race still appears, are recorded during these steps to guide the subsequent retries.
For packages with multiple pre-existing data races, it is crucial to determine whether the targeted data race has been fixed,
distinguishable via the stable bug hash. 

\subsubsection{Retries}

For attempts that fail to produce a fix, we implement a feedback loop that includes the failure message along with the rest of the prompt. 
This additional context helps the LLM generate a better fix to circumvent the encountered error.

For retries, we first attempt the function scope. If a validation error occurs, we proceed to the file scope without providing error feedback. If a validation error still occurs at the file scope, we retry once more, this time including diagnostic information in the feedback.

\subsection{Developer Validation}
For each race report with a validated fix, we automatically generate a patch and send it to the corresponding code owner(s) for review. 
If the patch is approved, it is automatically merged into the codebase. 
Rejected patches are reviewed by the \drfix{} development team for further improvement.

\section{Evaluation}
\label{sec:eval}

To empirically evaluate the effectiveness and usefulness of \drfix{} in producing fixes for data races, we conducted a series of experiments designed to answer the following research questions:

\begin{enumerate}
    \item \textbf{RQ1.} How effective is \drfix{} in generating fixes for data races?
    \item \textbf{RQ2.} What is the contribution of each component of the technique to the overall results?
    \item \textbf{RQ3.} How much do the results improve when utilizing a more advanced model?
    \item \textbf{RQ4.} How do developers perceive the usefulness of \drfix{} in their development process?
\end{enumerate}

\subsection{Experimental Setup.}
We used \company{}'s Go codebase for our evaluation.
As LLM models, we used GPT-4 Turbo~\citep{Newmodel93:online}, GPT-4o~\citep{HelloGPT26:online}, and o1-preview~\citep{Introduc16:online}, chosen due to their position as state-of-the-art models available during deployment of \drfix{}. 
We populated the vector database with \cureatedbugs{} previously fixed data race; the category of races and their frequencies are shown in Table~\ref{tab:category}.
On encountering a build or test failure during validation, we restricted to one retry with the failure fed back to the model.

Notably, there is no rule-based baseline for comparing \drfix{}'s LLM results on \company{}'s Go codebase, as no existing tools for fixing Go data races are available. To evaluate our approach, we conduct various ablation studies to empirically assess each component’s impact. We use developer acceptance rate relative to the number of proposed changes as a proxy for the quality of fixes in real-world scenarios, and passing test validation as a proxy for large-scale ablation studies.

\subsection{RQ1: How effective is \drfix{} in generating fixes for data races?} \label{subsec:rq1}

To evaluate the practical effectiveness of \drfix{}, we deployed it on the Go codebase at \company{}. 
We selected GPT-4 Turbo, which was the most advanced LLM model available at the time. 
Over a period of \drfixperiodnohyphern{}, we identified a total of 404 reproducible data race bugs in the Go codebase.

\drfix{} successfully generated automated fixes for 224 (55\%) of these data races. 
Among these generated fixes, developers approved 193 (86\%) for integration into the codebase. 
Eight of these approved fixes required minor idiomatic refinements before merging.
14\% of the proposed fixes were rejected by developers, primarily for reasons including prioritizing code readability over intricate solutions, opting for broader manual refactoring instead of targeted fixes, or identifying certain solutions as incorrect despite passing initial tests.
These observations highlight important opportunities for future improvements, such as expanding concurrency test coverage and incorporating direct developer feedback to enhance fix accuracy and usability.
Below we present a selection of real-world data races alongside their \drfix{} fixes, distilled to their key features. In practice, data race reports often span many stack frames across multiple files, complicating diagnosis. As Table~\ref{tab:survey_results} shows, even tasks that experts consider simple can challenge non-experts by requiring significant time to identify a race's root cause. \drfix{}'s automated repair significantly reduces this burden.

\begin{wraptable}{l}{0.6\textwidth}
\footnotesize
\caption{Data race category and their frequency in fixes produced by \drfix{} and in the vector database.}
    \centering
    \begin{tabular}{r|c|c}
\textbf{Category} &  \multicolumn{2}{c}{\textbf{Frequency}} \\
\ & \textbf{\drfix{} fixes} &  \textbf{VectorDB}\\
\midrule
Capture-by-reference in goroutines & 79 (41\%) & 102 (37.5\%)\\
Missing/incorrect synchronization & 50 (26\%) & 40 (14.7\%)\\
Parallel test suite & 26 (13\%) & 32 (11.8\%)\\
Capture of loop variable & 12 (6\%) & 7 (2.57\%)\\
Concurrent map access & 9 (5\%) & 14 (5.15\%))\\
Concurrent slice access & 9 (5\%) & 7 (2.57\%)\\
Others & 8 (4\%) & 70 (25.7\%)\\
\bottomrule    \end{tabular}
\label{tab:category}
\end{wraptable}
The minimum, average, median, and maximum durations to generate fixes were 6, 13, 14, and 29 minutes, respectively. Table~\ref{tab:category} provides a breakdown of the categories of data races fixed and their respective percentages.
We study in more detail why some data races remain unfixed, as part of \textbf{RQ2}.

\begin{figure}[!t]
  \begin{lstlisting}[language=CustomGo,label={casestudy:example4},caption=Example of Capture-by-reference in goroutines and the corresponding fix.,captionpos=b]
    ... // Rest of the code
limit := locationStoreConfig.Limit
    ... // Rest of the code
for _, params := range queryParams {
    go func(j entity.LatestPositionQueryParams) {
        defer wg.Done()
        now := c.clock.Now()
        end := now.Unix() * 1e3
        start := now.Add(-1*interval).Unix() * 1e3
@\gplus@       @\textcolor{darkgreen}{localLimit := limit}@
@\label{excby:ifthen}@        if entity.DiversityLevelEachProvider == *j.DiversityLevel {
@\rminus@           @\textcolor{red}{limit = locationStoreConfig.LimitForDiversityLevel}@
@\gplus@           @\textcolor{darkgreen}{localLimit = locationStoreConfig.LimitForDiversityLevel}@
        }
@\label{excby:req}@        request := entity.PositionQueryParams{
            EntityUUID:      j.JobUUID,
            StartTimeMillis: start,
            EndTimeMillis:   end,
@\rminus@           @\textcolor{red}{Limit:           limit,}@
@\gplus@           @\textcolor{darkgreen}{Limit:           localLimit,}@
            EntityType:      entity.LSEntityTypeFreightJob,
        }
    ... // Rest of the code (No more use of limit)
  \end{lstlisting}
\end{figure}

\textbf{Capture-by-reference in goroutines (41\%)} is the most prevalent data race category. These data races occur when goroutines unintentionally capture free variables from the outer scope.

Listing~\ref{casestudy:example4} shows an example where each goroutine created inside the for-loop captures the \mgo{limit} variable from the surrounding scope.
The if-then block on line~\ref{excby:ifthen} updates the value of \mgo{limit} with an intent to create a new \mgo{request} structure (line~\ref{excby:req}) with this updated value. However, the update to \mgo{limit} not only accidentally affects the parent scope, but may also cause goroutines to corrupt each other's data by overwriting the shared value of \mgo{limit}.

\drfix{} correctly addresses the race by introducing a local variable \mgo{localLimit}, ensuring that each goroutine operates on its copy of \mgo{limit}. 
This fix demonstrates the flexibility of \drfix{}, and GenAI in general, as it not only eliminates the data race via local variable declarations, but also correctly patches the program with \mgo{localLimit} at all subsequent locations within the goroutine.

\textbf{Missing or Incorrect Use of Synchronization (26\%)} is the second most common category, the underlying cause of which is developer misuse of synchronization mechanisms, for example, mutexes or atomics, when accessing shared resources.
Listing~\ref{casestudy:example1} depicts an incorrect placement of \mgo{Add} within a \mgo{WaitGroup} synchronization construct. 
Line~\ref{syncbug:define} defines a data structure, \mgo{proposal}, which includes a map and a lock (not shown). 
Next, it spawns 100 goroutines.
The parent goroutine uses a \mgo{sync.WaitGroup} to wait for the termination of all goroutines (line~\ref{syncbug:wait}) before proceeding.
Each goroutine calls \mgo{proposeReplica}, which safely updates the shared map within \mgo{proposal} by using the lock (\mgo{proposeReplica} internals are elided for brevity). 
However, \mgo{wg.Add} is incorrectly placed within the goroutine on line~\ref{syncbug:badadd}. 
Since \mgo{wg.Add} is executed only after the goroutine starts, it fails to increment the \mgo{wg} counter in time. 
Consequently, the \mgo{wg.Wait} on line~\ref{syncbug:wait} may observe a lower counter value than expected, potentially unblocking the parent before all children goroutines finish. 
The developer assumes exclusive access to the map \mgo{proposal.Proposal} after the \textbf{wg.Wait} statement and accesses it without any lock  (line~\ref{syncbug:access}).
But the children goroutines could be accessing the same map (albeit holding a lock) causing a data race between the parent and the children.

The correct approach is \emph{not} adding a lock after \mgo{wg.Wait}, but to move \mgo{wg.Add} before launching the goroutine, as shown in line~\ref{syncbug:correctadd}.
This ensures the parent goroutine observes an accurate counter value before waiting, which avoids it being unblocked prematurely. 
Notably, \drfix{} can address the data race without accessing the internal implementation of \mgo{proposeReplica}, or detailed knowledge of the \mgo{proposals} data structure.

\begin{figure}
  \begin{lstlisting}[language=CustomGo, label={casestudy:example1},caption=Example of incorrectly placing the \mgo{Add} method of a \mgo{WorkGroup} and the corresponding fix.,captionpos=b]
... // Rest of the code
// proposals contains a map (Proposal) and a lock (mutex) to protect it.
@\label{syncbug:define}@proposals := parser.NewKeyProposal()
var wg sync.WaitGroup
for i := 1; i < 100; i++ {
@\gplus@   @\label{syncbug:update}\textcolor{darkgreen}{wg.Add(1)}@
    go func (pod int) {
@\rminus@       @\label{syncbug:badadd}\textcolor{red}{wg.Add(1)}@
        defer wg.Done()
        // updates the map proposals.Proposal holding the proposals.mutex.
        @\label{syncbug:correctadd}@ proposeReplica(proposals, i)
    }(i)
}
@\label{syncbug:wait}@ wg.Wait() // Wait ends prematurely
@\label{syncbug:access}@ if for k := range proposals.Proposal {...} // Accesses the Proposal map without locking
... // Rest of the code
  \end{lstlisting}
\end{figure}

\begin{wrapfigure}{l}{0.5\textwidth}
  \begin{lstlisting}[language=CustomGo,label={casestudy:example3},caption=Shared struct fields in a parallel test suite resulting a data race and the corresponding fix.,captionpos=b]
func TestUploadReaderRead(t *testing.T) {
  sampleReader := strings.NewReader("hi")
  sampleReader2 := strings.NewReader("hi")
@\rminus@ @\textcolor{red}{ sampleHash := md5.New()}@
  tests := []struct {
    name    string
    object       io.Reader
    hash         hash.Hash}{
    { name:   "Success - 1",
      object:  sampleReader,
@\rminus@     @\textcolor{red}{ hash:\ \ \ \ \ \ sampleHash,}@ 
@\gplus@     @\textcolor{darkgreen}{ hash:\ \ \ \ \ \ md5.New(),}@
      },
    { name:    "Success - 2",
      object:  sampleReader2,
@\rminus@     @\textcolor{red}{hash:\ \ \ \ \  \ sampleHash,}@ 
@\gplus@     @\textcolor{darkgreen}{hash:\ \ \ \ \ \ md5.New(),}@
      },
  }
  for _, tt := range tests {
    tt := tt
    t.Run(tt.name, func(t *testing.T) {
      t.Parallel()
      ur := UploadReader{
          object:       tt.object,
          hash:         tt.hash,
        }
        gotN, err := ur.Read(tt.args.p)
          ... // Rest of the test
        })
    }
}
  \end{lstlisting}
\end{wrapfigure}
\noindent

\textbf{Parallel Test Suite (Table-Driven Testing) 
(13\%)} is another significant category, in which data races occur whenever tests running in parallel share mutable state without proper isolation.
Listing~\ref{casestudy:example3} illustrates such a case. 
The function \mgo{TestUploadReaderRead} tests the reading of the data using an \mgo{io.Reader}, which is then hashed using the MD5 hash function. 
The test concurrently runs multiple sub-tests, each representing a test case, by using \mgo{t.Parallel}. 
In this example, the data race is caused by unsafe access of the shared object \mgo{sampleHash} among the sub-tests. 
Multiple test cases attempt to concurrently write to the same \mgo{hash.Hash} inside the \mgo{Read} method, leading to unpredictable results and potential data corruption.
We note that the racing source lines are in the code under test (not shown), but the root cause is in the test code.

To fix the data race, \drfix{} replaces the shared object with new, independent instances in each test case.
This is achieved by removing the \mgo{sampleHash} variable defined in the parent scope, and instead creating a new instance of the hash function with \mgo{md5.New} for each test case.

This privatization ensures that each parallel test operates on its own independent objects, avoiding the data race.
This fix illustrates how \drfix{} accurately identifies the lack of private copies in parallel tests as the main issue and responds by inserting as many copies and deleting the shared object.
This example also shows the effectiveness of \drfix{} to come up with a fix by exploring different scopes --- here the fix is not in the leaf functions where the racing accesses occur.

\begin{figure}
    \centering
    \begin{minipage}[b]{0.48\textwidth}
        \begin{lstlisting}[language=CustomGo, label={casestudy:example6}, caption=Data race due to concurrent access to the built-in thread-unsafe map and its fix via \mgo{sync.Map}., captionpos=b]
type Scanner[ROW any] struct {
@\rminus@  @\label{map:old}\textcolor{red}{ lockMap  map[sharding.ShardKey]shardLock}@
@\gplus@  @\label{map:sync}\textcolor{darkgreen}{ lockMap    sync.Map}@
}
func (t *Scanner[ROW]) runForever() {
  for {
    ... // Rest of the code
    go t.runShards()
    ... // Rest of the code
   }
}
func (t *Scanner[ROW]) runShards() {
  newShards := t.shardCtrl.MyShards()
@\rminus@  @\label{map:for}\textcolor{red}{ for removedShard := range t.lockMap \{}@
@\rminus@    @\textcolor{red}{ if \_, ok := newShards[removedShard];!ok \{}@
@\rminus@      @\label{map:delete}\textcolor{red}{ delete(t.lockMap, removedShard)}@
@\rminus@    @\textcolor{red}{ \}}@
@\rminus@  @\textcolor{red}{ \}}@
@\gplus@  @\label{map:range}\textcolor{darkgreen}{ t.lockMap.Range(func(key,value interface{}) bool \{}@
@\gplus@    @\textcolor{darkgreen}{ shardKey := key.(sharding.ShardKey)}@
@\gplus@    @\textcolor{darkgreen}{ if \_, ok := newShards[shardKey]; !ok \{}@
@\gplus@      @\label{map:syndel}\textcolor{darkgreen}{ t.lockMap.Delete(shardKey)}@
@\gplus@    @\textcolor{darkgreen}{ \}}@
@\gplus@    @\textcolor{darkgreen}{ return true}@
@\gplus@  @\textcolor{darkgreen}{ \})}@
  ... // Rest of the code
}
        \end{lstlisting}
    \end{minipage}
    \hfill
    \begin{minipage}[b]{0.48\textwidth}
        \centering
  \begin{lstlisting}[language=CustomGo,label={casestudy:example7},caption=Data race due to concurrent concurrent slice access and its fix via locks.,captionpos=b]
type fmrImpl struct {
  ...
  ch    []chan struct{}
@\label{slice:mutex}@@\gplus@ @\textcolor{darkgreen}{ chLock   sync.Mutex}@
}

func (fmr * fmrImpl) Update(...) {
  ...
@\gplus@ @\textcolor{darkgreen}{ fmr.chLock.Lock()}@
  newChan := make(chan struct{})
@\label{slice:append}@  fmr.ch = append(fmr.ch, newChan)
@\gplus@ @\textcolor{darkgreen}{ fmr.chLock.Unlock()}@
  ... // Rest of the code
}
func (fmr * fmrImpl) setSync(idx int32) {
  go func() {
    ...
    select {
      ...
@\rminus@     @\label{slice:read1}\textcolor{red}{case <-fmr.ch[idx]:}@
@\gplus@     @\textcolor{darkgreen}{ case <-func\(\) chan struct\{\} \{}@
@\gplus@        @\textcolor{darkgreen}{ fmr.chLock.Lock()}@
@\gplus@        @\textcolor{darkgreen}{ defer fmr.chLock.Unlock()}@
@\gplus@        @\label{slice:read2}\textcolor{darkgreen}{ return fmr.ch[idx]}@
@\gplus@     @\textcolor{darkgreen}{ \}():}@
      return
    }
  }()
}
  \end{lstlisting}
    \end{minipage}
\end{figure}

\textbf{Concurrent Map Access (5\%)} occur because maps in Go are not thread-safe by default, i.e., concurrent reads and writes without synchronization may lead to data races.

Listing \ref{casestudy:example6} shows an example, where the \mgo{runForever} method of the \mgo{Scanner} object periodically creates a goroutine that executes \mgo{runShards}. 
Each goroutine processes shard-related data by cleaning up obsolete shard locks stored in \mgo{lockMap}. 
Since multiple instances of \mgo{runShards} execute concurrently, there is a risk of data races when they access or modify the shared \mgo{lockMap} concurrently.
Even though the shards stored in the maps are lock protected, the map itself is not.

\drfix{} addresses the data race by replacing the standard map with a \mgo{sync.Map}, a thread-safe map implementation specifically designed for concurrent access.
This fix demonstrates how \drfix{} identifies that \mgo{lockMap} (line~\ref{map:old}) is not thread-safe, and replaces it with the thread-safe \mgo{sync.Map} version (line~\ref{map:sync}).
Furthermore, since \mgo{sync.Map} cannot be iterated via a for-loop unlike the built-in \mgo{map} (line~\ref{map:for}), \drfix{} correctly substitutes the loop with a \mgo{Range} API and constructs the anonymous function parameter required by \mgo{Range} (line~\ref{map:range}).
\drfix{} also replaces the built-in \mgo{delete} function (line~\ref{map:delete}), which only operates on built-in \mgo{map} types, with a call to the \mgo{Delete} API of \mgo{sync.Map} (line~\ref{map:syndel}) to accomplish the same objective.

\textbf{Concurrent Slice Access (5\%)} occur because slices, like maps, are not safe for concurrent access. Modifying slices from multiple goroutines without synchronization can cause race conditions. Listing~\ref{casestudy:example7} shows an example of a code snippet that manages periodic synchronization tasks and controls goroutine lifecycle using channels. 
The \mgo{ch} is slice of channels. 
In \mgo{Update}, a goroutine appends new channels to the \mgo{ch} slice on line~\ref{slice:append}. 
In \mgo{setPeriodicSync}, another goroutine concurrently reads from a specific index of \mgo{ch} in its select-case at line~\ref{slice:read1}. 
Concurrent reads from \mgo{ch} while its size is being changed causes a data race. 
\drfix{} fixes it by introducing a mutex \mgo{sync.Mutex} named \mgo{chLock} (line~\ref{slice:mutex}) in the \mgo{fmrImpl}. 
The slice is lock protected both at the \mgo{append} site (line~\ref{slice:append}) and at indexing site (line ~\ref{slice:read2}). 
Interestingly, the fix in the \mgo{setPeriodicSync} function uses a lambda function that wraps the indexing of \mgo{ch}, protecting it with the common lock and uses the \mgo{defer} statement, making it an idiomatic fix.

\textbf{Capture-by-Reference of Loop Variable (6\%)} arises when loop variables are captured by reference in closures, causing all closures to reference the same variable instance. Listing~\ref{casestudy:exampleCaptureByReference} in Appendix~\ref{sec:loopvarcapture} shows an example. Due to the notoriety of this race, go 1.22 changed the for-loop semantics to make range variables have per-iteration scope instead of per-loop scope~\citep{go1.22:online}.

\textbf{Others (4\%)} includes miscellaneous data race categories not covered above. For instance, a handler function generates random bytes for HTTP responses using the Go \mgo{math/rand} package. A race occurs when multiple requests concurrently access a shared \mgo{rand.Source}, which is not safe for concurrent use. This is resolved by creating a new \mgo{rand.Source} per request shown in Appendix~\ref{sec:miscrand}.

\subsection{RQ2: What is the contribution of each component to the overall results?}

We conducted an ablation study to evaluate the contribution of each component in our technique to the fixes generated by \drfix{}. 
We initially collected \cureatedsupersetbugs{} data race bugs detected and fixed in the Go codebase over the past few years, providing ground truth information for evaluating \drfix{}’s success on real-world issues. 
Of these, we successfully reproduced the data races in 403 (54\%) cases. 
The remaining data races could not be reproduced due to significant changes to the codebase. \drfix{} was then applied to these 403 reproducible data races.
These 403 examples are distinct from the 272
examples used to prepare the vector database used for the RAG mentioned in Section~\ref{sec:example}.

For the study, GPT-4o was chosen as LLM due to its availability at the time of this evaluation. 
To assess their impact of various components within \drfix{}, we selectively toggled specific features. 
Importantly, developer acceptance was not evaluated in this ablation study, as it would have introduced extensive back-and-forth processes beyond the scope of the automated setup.
Instead, we relied on test validation and spot checking to achieve large-scale results. 
While this approach may overcount the impact of certain components, it provides faithful comparisons across experiments, since the same strategy is consistently applied with and without the feature under study.

In this evaluation, we would like to answer the following questions.
\begin{description}
    \item[RQ2.1] How effective are example bugs and their corresponding fixes in guiding LLMs to produce fixes,  given that LLMs are already trained on a vast text corpus?
    \item[RQ2.2]  How effective is concurrency structure-based example selection (skeletonization) in producing higher-quality fixes?
    \item[RQ2.3]  \label{it:scope} Does the choice of \emph{fix scope} (functions vs. files) and their order make a measurable impact?
    \item [RQ2.4] Does providing the feedback of validation failure improve the results?    
    \item [RQ2.5] \label{it:loc} Does the choice of \emph{fix location} (leaf, test, LCA) and their orders have a measurable impact?    
\end{description}

\paragraph{Impact of examples (RAG) on LLM} Figure~\ref{fig:rag} presents the results of an ablation study that answers \textbf{RQ2.1} and \textbf{RQ2.2}.
We compare three setups (1) \textbf{No RAG}, where we ask the LLM to fix the defect without providing any example bug/fix; (2) \textbf{RAG without skeleton}, where we retrieve an example via standard textual similarity, and (3) \textbf{RAG with skeleton}, where we retrieve a prior fix by matching the concurrency skeleton as described in Section~\ref{sec:match-and-fix}.

\begin{figure}[t!]
    \centering
    \begin{minipage}[b]{0.49\textwidth}
        \centering
        \includegraphics[width=.9\linewidth]{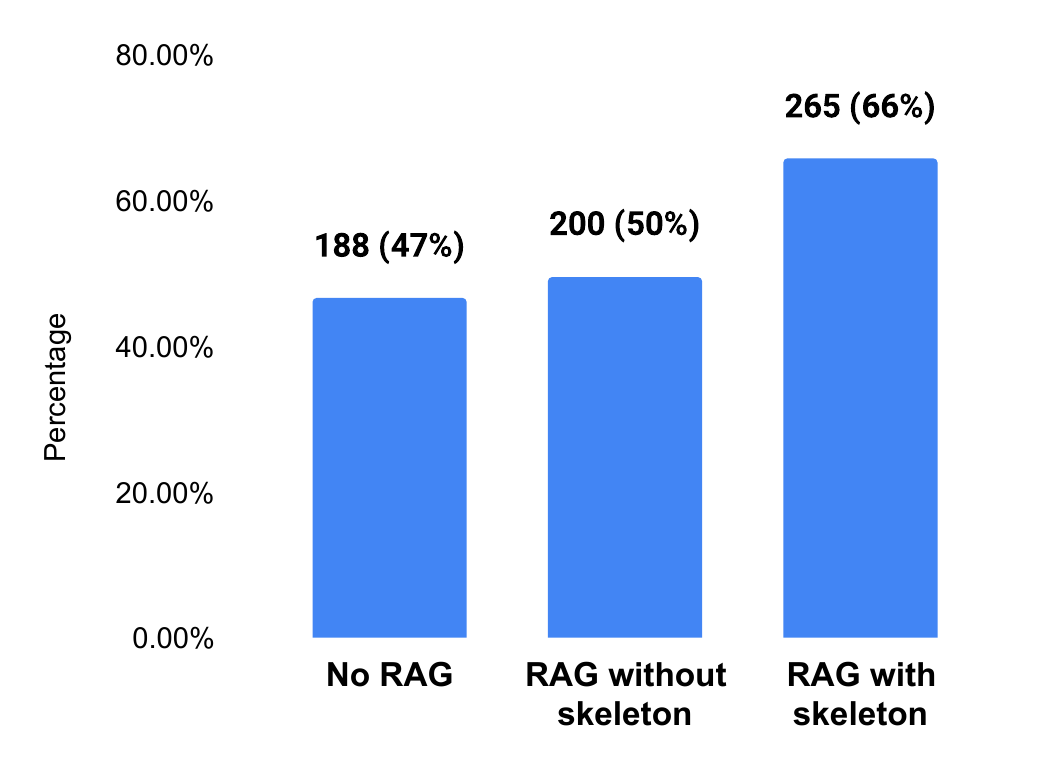}
        \caption{Comparison of how using examples and selecting them via concurrency skeleton impacts the number of successful fixes.}
        \label{fig:rag}
    \end{minipage}
    \hfill
    \begin{minipage}[b]{0.49\textwidth}
        \centering
        \includegraphics[width=.9\linewidth]{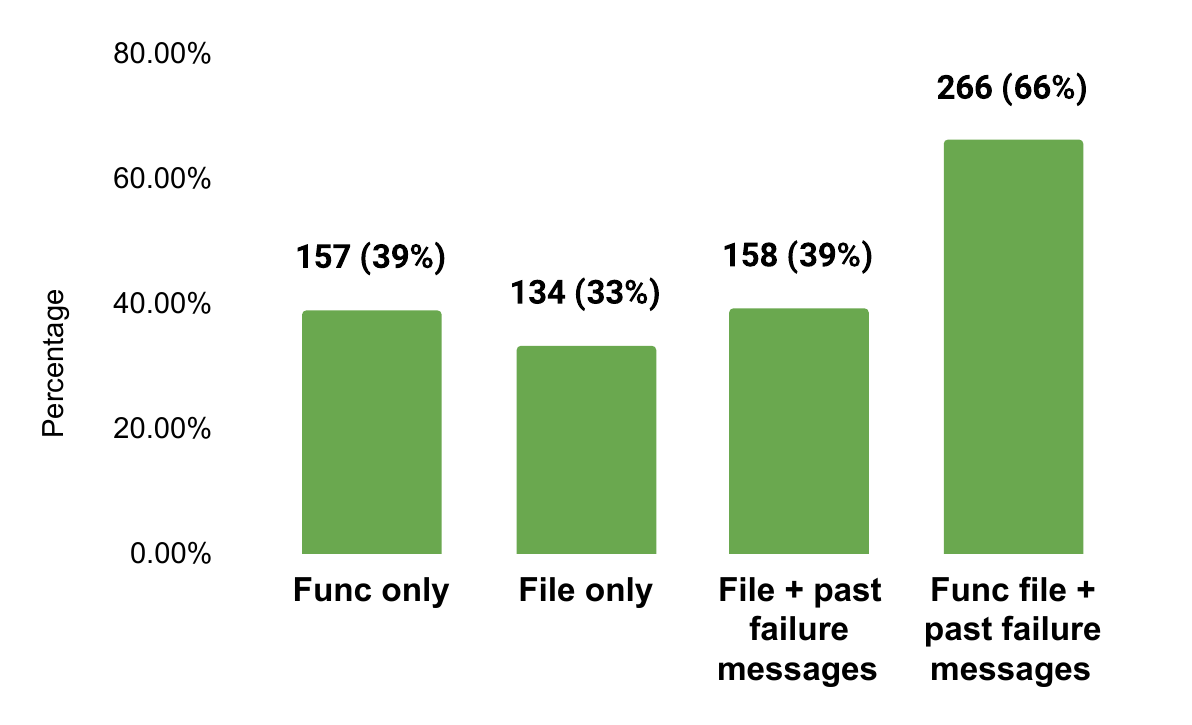}
        \caption{Comparison of how fixing at different scopes and using the diagnostics from previous failures improves the success rate.}
        \label{fig:funcfile}
    \end{minipage}
\end{figure}

\drfix{} successfully generated fixes for 47\%, 50\%, and 66\% of the data races in ``No RAG'', ``RAG without skeleton'', and ``RAG with skeleton experiments'', respectively. 
The results indicate that using RAG without skeleton slightly increases (about 3\%) the number of validated fixes. 
However, using RAG with skeleton demonstrates approximately a 19\% improvement in the number of validated fixes. 
This significant improvement confirms the effectiveness of our novel design of code skeletonization. 
Listing~\ref{casestudy:example6} is an example of a fix generated by \drfix{} using RAG with skeleton.

Empirical evidence shows that some data races remain unfixed when our LLM is prompted without retrieval-augmented generation (RAG), yet the same races are successfully patched once RAG is enabled. 
A study of developer-approved fixes that \drfix{} could only generate through RAG revealed that in most of these cases either:

\begin{itemize}
  \item The code exhibited \emph{complex control and concurrency} constructs such as mixed use of shared-variables and message passing or mixing message passing with workgroups/locks or
  \item The solution demanded introducing complex control and concurrency constructs and/or \emph{restructuring} code across multiple code blocks.
\end{itemize}
In these scenarios, the plain LLM alone failed to incorporate the necessary reorganization of code. 
Once guided by RAG, however, the model accessed structurally similar (previously patched) data races from the curated knowledge base. 
These examples effectively narrow the LLM’s search space and highlight concurrency primitives---\texttt{sync.Map}, read-write locks, privatization, or explicit atomic operations---that have succeeded in analogous contexts. 
As a result, the LLM is ``nudged'' into assembling more complex solutions in a refined space, which it had failed to discover on its own. Table~\ref{tab:complexrags} shows the recurring complex patterns in fixes where the RAG played a pivotal role.

\begin{table}[!t]
\scriptsize
    \centering
    \caption{Fixes where RAG played a pivotal role.}
\begin{tabular}{c|l}
    1 & Creating copies of complex data structures (e.g., nested fields in structs or slices) to avoid unwanted sharing. \\ \hline
    2 &  Changing function signatures to pass additional parameters and avoid accidental reference capture, \\ 
    & along with updates to all associated call sites.  \\ \hline
    3 &  Changing data types (value vs. pointer; \texttt{map} vs. \texttt{sync.Map}) and propagating those changes to all references. \\     \hline
    4 &  Introducing atomic operations on shared variables (including nested fields) and carefully  \\
    & (e.g., type casting or using the appropriate atomic API for the type) replacing reads with \texttt{atomic.Load}s,\\
    & writes with \texttt{atomic.Store}/\texttt{CompareAndSwap}/\texttt{Add}. \\ \hline
    5 &  Introducing a new mutex into a larger aggregate type,\\
    & along with lock/unlock calls that precisely guard all usage points of multiple different shared data. \\ \hline
    6 &  Managing multiple locks in different code regions to fix a single data race, which often requires carefully scoping each lock. \\ \hline
    7 &  Switching to reader-writer locks and upgrading/downgrading them in different places, \\
    & including changing any type declarations involved. \\ \hline
    8 &  Adding, deleting, or relocating synchronization constructs such as \texttt{Add}/\texttt{Done}/\texttt{Wait} on a \texttt{WaitGroup}, \\ 
    & or appropriately placing \texttt{send}/\texttt{recv} in channels and \texttt{select} statements. \\ \hline
    9 &  Dealing with multi-level goroutine nesting. \\
    \end{tabular}
    \label{tab:complexrags}
\end{table}

Using the RAG also eliminates repeated prompting of the base LLM at varying temperatures to “guess” a correct fix. 
Instead, \drfix{} retrieves structurally aligned solutions from a vetted store of concurrency patches, drastically reducing random attempts and increasing the chance of an immediately valid fix.
Appendix~\ref{sec:rag-example-fix} in the supplemental material lists many RAG examples.

\paragraph{Impact of fixing scope, their order, and validation feedback}
To answer \textbf{RQ2.3} and \textbf{RQ2.4},  we conducted experiments with these settings using RAG with skeletons
(1) fix in function-only scope, (2) fix in file-only scope, (3) fix in file-only scope with  past validation failure (if any) fed back to the LLM, and (4) fix function-then-file scope with validation failure (if any) fed back to the LLM.
The last choice represents how we do it in our production settings.
As illustrated in Figure~\ref{fig:funcfile}, \drfix{} generated fixes for 39\%, 33\%, 39\%, and 66\% of data races across these configurations, respectively. 
These results suggest that the function-only context can be succinct but limited, while file-only context is comprehensive yet possibly overwhelming for the LLM. 
The function-then-file with validation failure fed back from previous steps leverages the advantages of succinct and comprehensive information, achieving optimal results by combining both scope types and including failure information.

Modern LLMs' expanded context windows might suggest that providing the broadest possible context---such as entire bug reports with all relevant file contents---at once would produce accurate fixes. However, our studies indicate that this approach actually reduces success rates. This outcome aligns with known limitations of transformer-based models, which struggle to consistently maintain attention across long contexts~\citep{liu2024lost}.
Our approach of scoping the context to likely solution spaces offers a much higher success rate.

\paragraph{Impact of fix location and their order} 
Table~\ref{tab:category} already shows that 15\% of our fixes target test files, which shows the impact of the leaf-function vs. the test function in the call stack.
To assess the role of the Lowest Common Ancestor (LCA) fix type, we performed two experiments: (1) Without LCA and (2) With LCA. 
\drfix{} generated fixes for 62.53\% of cases without LCA and 66.75\% with LCA, showing a 4\% improvement from adding LCA, which highlights its modest but meaningful impact.

\paragraph{Where \drfix{} fails to produce a fix:}
To gain further insights into unsuccessful cases, we manually reviewed developer solutions, categorizing them to pinpoint \drfix{} limitations and highlight areas for future improvements. Table~\ref{tab:data_race_categories} presents a breakdown of these categories, each accompanied by a brief description. 
Our analysis showed that common reasons for failures include tool constraints, notably the current restriction to analyzing two files, and situations requiring significant code refactoring. Additionally, removing parallelism emerged as a frequent manual solution that \drfix{} did not suggest, as we intentionally avoided altering the programmer's original intent.

\begin{table}[t!]
    \centering
    \scriptsize
    \caption{Categories of data races not fixed by \drfix{}}
    \label{tab:data_race_categories}
    \begin{tabular}{
        >{\raggedright\arraybackslash}p{4.5cm}  
        >{\centering\arraybackslash}p{2cm}      
        >{\raggedright\arraybackslash}p{6cm}    
    }
        \toprule
        \textbf{Category} & \textbf{Frequency (\%)} & \textbf{Short Description} \\
        \midrule
        More than 2 File Changes & 29 (21\%) & Requires changes across multiple files, complicating automated fixes. \\ \hline
        
        Change/Reduce/Remove Parallelism & 27 (19\%) & Involves altering concurrency, potentially changing intended behavior. \\ \hline
        
        Change the Business Logic & 21 (15\%) & Needs adjustments to core functionality, which tools avoid to prevent side effects. \\ \hline
        
        Unable to Isolate the Failing Test & 14 (10\%) & Large tests make pinpointing the data race source difficult. \\ \hline
        
        External & 14 (10\%) & Issues in external code that cannot be modified by the tool. \\ \hline
        
        Large Code Refactoring & 8 (6\%) & Requires extensive code changes beyond automated fixing capabilities. \\ \hline
        
        Others & 8 (6\%) & Miscellaneous cases with unique challenges for automated fixes. \\ \hline
        
        Using Deep Copy & 7 (5\%) & Fixing requires deep copying objects, complex for automated tools. \\ \hline
        
        Singleton Pattern & 5 (4\%) & Shared instances cause races, needing redesign of instance management. \\ \hline
        
        Non-trivial Even for Experts & 5 (4\%) & Complex data races challenging even for experienced developers. \\
        \bottomrule
    \end{tabular}
\end{table}

\subsection{RQ3: How much do the results improve when utilizing a more advanced model?}\label{subsec:advanced_model}

We now focus on the impact of more advanced LLMs.
As LLMs continue to improve, we assessed their evolving capabilities by comparing the performance of two models: GPT-4o, released on May 13, 2024, and GPT-o1-preview, released on September 12, 2024. 
We used the same set of 403 reproducible data race bugs from our previous analysis (RQ2). With GPT-4o, \drfix{} generated fixes for 65.76\% of these bugs, whereas with GPT-o1-preview, the fix rate increased to 73.45\%, resulting in a 7.7 percentage point improvement over the baseline.

Listing~\ref{casestudy:example10} shows an example of a data race fixed by the GPT-o1-preview model, but not GPT-4o. 
The issue is subtle.
The data race happens when concurrently writing to the \mgo{err} variable on line~\ref{4o:race1} while reading from line~\ref{4o:race2}.
The parent goroutine \mgo{EvaluateRisk} spawns the child goroutine \mgo{run} at line~\ref{4o:go}.
The \mgo{run} goroutine calls \mgo{c.Evaluate}, whose results are written to \mgo{result} and \mgo{err}.
\mgo{err} is captured by reference from the parent scope.

\begin{wrapfigure}{l}{0.5\textwidth}
    \centering
  \begin{lstlisting}[language=CustomGo,label={casestudy:example10},caption=An example fixed by the newer o1-preview model that was not fixed by the older GPT-4o model.,captionpos=b]
func (c *controller) EvaluateRisk(
    ctx context.Context, req *Request)
    (*Response, error) {
    resultChan := make(chan Result, 1) 
    var resp *Response 
@\rminus@   @\textcolor{red}{var err        error}@
@\label{4o:errChan}\gplus@   @\textcolor{darkgreen}{errChan    = make(chan error, 1)}@
    run := func() {
@\label{4o:resultVar}\rminus@         @\textcolor{red}{ var result Result}@
@\label{4o:race1}\rminus@         @\textcolor{red}{ result, err = c.Evaluate(ctx, req)}@
@\gplus@         @\textcolor{darkgreen}{ result, err := c.Evaluate(ctx, req)}@
          resultChan <- result
@\label{4o:childSend}\gplus@         @\textcolor{darkgreen}{ errChan <- err }@
    }
@\label{4o:go}@    go run()
@\gplus@   @\textcolor{darkgreen}{var err error }@
@\label{4o:select}@    select {
@\label{4o:readResultChan}@        case result := <-resultChan:
@\label{4o:readErrCh}\gplus@           @\textcolor{darkgreen}{ err = <-errChan }@
            ...
@\label{4o:done}@        case <-ctx.Done():
    ... // Rest of the code
@\label{4o:race2}@    return resp, err
}
  \end{lstlisting}
\end{wrapfigure}

The \mgo{result} is a goroutine-local variable defined on line~\ref{4o:resultVar}.
The child goroutine uses the channel \mgo{resultChan} to send the result to the parent waiting on the \mgo{select} statement on line~\ref{4o:select}.
In normal execution, when the child finishes and sends its results over \mgo{resultChan}, it establishes a happens-before relation between the write to the \mgo{err} on line~\ref{4o:race1} and its subsequent read in the parent on lines~\ref{4o:race2}.

However, if the \mgo{ctx} context, which holds execution deadlines, expires, the select statement takes the \mgo{case <- ctx.Done()} arm (line~\ref{4o:done}) without waiting for the child goroutine to finish. 
In this execution, when the parent goroutine reads from \mgo{err} on line~\ref{4o:race2}, the child could be updating it on line~\ref{4o:race1}, without a happens-before ordering, causing the data race.

To fix the data race, \drfix{} removes the sharing of \mgo{err} between the goroutines; instead it introduces another channel \mgo{errChan} (line~\ref{4o:errChan}) to send the error value from the child to the parent (line~\ref{4o:childSend}).
The parent reads from \mgo{errChan} on line~\ref{4o:readErrCh}, assigning its value to the local \mgo{err} variable.
Now, if the context expires, the \mgo{err} variable is not concurrently modified.
Furthermore, both channels are buffered ensuring no accidental partial deadlock~\citep{saioc2024unveiling, saioc2025dynamic}, a different concurrency bug.

\subsection{RQ4: How do developers perceive \drfix{} in their development workflow?}

\begin{table}
    \centering
    \scriptsize
    \caption{Survey Results on Developers' Perceptions of \drfix{}}
    \label{tab:survey_results}
    \begin{minipage}{0.4\linewidth}
        \centering
        \begin{tabular}{ll}    
        \toprule
        \multicolumn{2}{c}{\textbf{Go Programming Experience}} \\
        \midrule
        \textbf{Experience Level} & \textbf{Count(\%)} \\
        \midrule
        Less than 1 year          & 5  (24\%) \\
        1 to 3 years              & 9  (43\%) \\
        3 to 5 years              & 3  (14\%) \\
        More than 5 years         & 4  (19\%) \\
        \midrule
        \multicolumn{2}{c}{\textbf{Familiarity with Concurrency in Go}} \\
        \midrule
        \textbf{Familiarity Level} & \textbf{Count(\%)} \\
        \midrule
        Somewhat Familiar           & 12 (57\%) \\
        Very Familiar               & 9  (43\%) \\
        \midrule
        \textbf{Total Developers} & 21  \\
        \bottomrule
    \end{tabular}
    \end{minipage}
    \hfill
    \begin{minipage}{0.55\linewidth}
        \centering
        \begin{tabular}{ll}
            \toprule
        \midrule
        \multicolumn{2}{c}{\textbf{Comfort Level in Fixing Data Races}} \\
        \midrule
        \textbf{Comfort Level} & \textbf{Count(\%)} \\
        \midrule
        Not Comfortable at All                & 1  (5\%) \\
        Slightly Comfortable but Need Help    & 14 (67\%) \\
        Very Comfortable and Do Not Need Help & 6  (28\%) \\
        \midrule
        \multicolumn{2}{c}{\textbf{Quality and Complexity of Data Races fixed by \drfix{}}} \\
        \midrule
        \textbf{Metric} & \textbf{Average Rating (stddev)} \\
        \midrule
        Quality of Fixes (1-5)               & 3.38 $\pm$1.24 \\
        Complexity of Data Races (1-5)       & 3.00 $\pm$0.89 \\
        \midrule
        \multicolumn{2}{c}{\textbf{Estimated Time Saved by Using Dr.Fix}} \\
        \midrule
        \textbf{Time Saved} & \textbf{Count(\%)} \\
        \midrule
        Up to 1 day                & 14 (67\%) \\
        1 to 2 days               & 4  (19\%) \\
        2 to 4 days               & 2  (9\%)  \\
        1 to 2 weeks              & 1  (5\%)  \\
        \bottomrule
    \end{tabular}
    \end{minipage}
\end{table}

We surveyed developers to assess their perception of \drfix{} and better understand how it enhances the development process, with 21 responses.
Table~\ref{tab:survey_results} shows the results, which indicate that developers have varying levels of experience and expertise in Go programming and concurrency concepts; and \drfix{} can generate fixes of satisfactory quality for data races of moderate complexity. 
\drfix{} achieved a satisfaction score of 3.38 out of 5 (1 being low and 5 being high), equating to 67.6\%. 
This result indicates a positive reception, as scores above 50\% are generally considered favorable. 

\begin{wraptable}{l}{0.5\textwidth}
\scriptsize
\centering
\caption{LoC comparison (Human vs. Dr.Fix).}
\label{tab:complexity}
\begin{tabular}{l|c|c|c||c}
    \toprule
    \textbf{\%tile} & \textbf{Human(H)} & \textbf{\drfix{}(D)} & \textbf{H/D} & \textbf{VectorDB} \\
    \midrule
    P50  & 10 &  9 & 1.11 &  7  \\
    P75  & 15 & 15 & 1.47 & 16  \\
    P90  & 46 & 29 & 1.59 & 29  \\
    P95  & 49 & 41 & 1.20 & 40  \\
    P99  & 97 & 46 & 2.11 & 71  \\
    P100 & 98 & 46 & 2.13 & 94  \\
    \bottomrule
\end{tabular}
\end{wraptable}

Additionally, estimated time savings indicate that \drfix{} helped streamline the debugging and resolution process. 
Empirically, we found that data race tickets historically took an average of 11 days from in-progress to closure, whereas \drfix{}-generated fixes reduced this timeline to three days.

The developers rated the average complexity of fixes produced by \drfix{} to be \(3.00 \pm 0.89\) (where \(1 = \text{simple}\) and \(5 = \text{complex}\)), 
indicating that \drfix{} handles non-trivial cases. 
Table~\ref{tab:complexity} shows the percentile spread of lines of code (LoC) changed in fixing data races by humans vs. fixes generated by \drfix{}; for reference the table also includes the LoC for the examples in the vector database.
Using the LoC changed to fix a data race as a proxy for the complexity of the defect, \drfix{}'s best fixes match the 90\textsuperscript{th} percentile (46 LoC) of human fixes.
Finally, the 193 fixes generated by \drfix{} add up to producing 2.1K lines of fixes.

\subsection{Limitations and Threat to Validity}
Not all incoming race reports traverse the entire pipeline. 
For instance, due to practical issues such as truncated calling contexts, we may be unable to determine the tests of which package produced the race report. 
We may also encounter difficulties in reproducing data races due to non-determinism, or whenever they have been fixed by a developer who, then, neglected to update the bug report. 
In other cases, we cannot reproduce old bugs due to infrastructure changes, upgrades, or migrations that prevent us from reverting to an earlier state when the bug was present. 
Often, the leaf racing statements involve external code, which does not fit within our internal developer workflows. 
In all such scenarios, we do not produce any candidate fix.

The current implementation of data race reports in Go by ThreadSanitizer limits the goroutine ancestry to only two levels. \drfix{} operates within this limitation, which is not fundamental and can be adjusted if more ancestry levels are supported.

The validator passing all tests with no data race detected is necessary, but not a sufficient condition to fix a data race.
But based on our experience, however, repeating \testrepeats{} times with the race detector is a reliable indicator. 
We plan to leverage additional techniques~\citep{sen2008race, samak2015synthesizing, cadar2008klee} to improve this further in future work.

We did not extensively evaluate open-source models such as Llama 3~\citep{dubey2024llama} and Code Llama~\citep{roziere2023code}, as their preliminary results were unpromising. 
We explored fine-tuning~\citep{raffel2020exploring} to train LLMs for data races,  but their results were unsatisfactory.
For that reason, we stick with the aforementioned Retrieval-Augmented Generation (RAG) strategy, which  is corroborated by other studies~\citep{ovadia2023fine}.

\subsection{Discussion}
We evaluated \drfix{} in an industrial Go environment, yet key questions remain about its broader applicability and effectiveness in other settings.

\subsubsection{Ability of \drfix{} to Fix Data Races Absent From the VectorDB}
RAG can fail if the retrieved examples differ too greatly from the current bug. Hence, \drfix{} first attempts a fix without RAG, then activates it only if needed. Even without training on proprietary code, our base model achieved a 47\% fix rate, rising to 66\% with RAG---demonstrating how curated examples substantially aid concurrency repair. Upgrading the LLM from \texttt{4o} to \texttt{o1} (without updating the VectorDB) also produced novel fixes (Section~\ref{subsec:advanced_model}), showing that races absent from the VectorDB can still be solved.

Nevertheless, the curated example set is finite. When neither the base model nor any retrieved example suffices, \drfix{} may fail. 
Our analysis shows that such failures often require beyond-two-file changes, removal of concurrency, or domain-specific refactoring absent from the VectorDB.

Code complexity is theoretically unbounded; hence, one cannot cover all possible concurrency scenarios. In this aspect, \drfix{} took a pragmatic stance over theoretical coverage.
Future research may investigate how VectorDB size relates to fix quality, the minimum sample set needed for improvement, the possibility of diminishing returns beyond a certain database size, and how best to balance widely encountered bugs against rarer ones. Such studies could help refine the equilibrium between curated examples and large language models in automated concurrency repair.

\subsubsection{Generality and Adaptability to Other Languages}
\drfix{}’s methodology is inherently adaptable. 
With a custom skeleton extraction tailored to a target language, a dedicated VectorDB of past fixes, and a frontend that accommodates different race report formats, \drfix{} can be extended beyond Go. 
Accurate source code location information remains crucial for effective analysis. 
Although Go was chosen for its microservice prevalence and high race susceptibility, we posit that \drfix{} can be applied to other languages---with effectiveness varying according to the quality of the available race detection tool and the quality of curated examples.

Foundational language models, such as the ones used in this work, are known to be adaptable to general tasks, and we expect the knowledge of fixing data races trained from a worldwide web of information to be transferable across languages. 
However, the extent to which an LLM trained on such information can successfully address data races in significantly different or previously unseen programming languages remains an open research question~\citep{LotfiNonVac}. Determining the efficacy of LLMs in adapting to distinct linguistic and concurrency patterns warrants a comprehensive empirical evaluation, which we propose as an important avenue for future research.

\subsubsection{Alternative sources of race detection}
As implemented, \drfix{} relies on a frontend that reads race reports containing detailed stack traces from dynamic analysis. 
Adapting the system to work with static-analysis tools---which may produce false positives and may lack precise interleaving details---would require modifications to the input format. 
While the absence of precise context might hinder fix localization, a static tool capable of pinpointing interleavings could, conversely, narrow the exploration space and guide the LLM toward a more precise solution. Evaluating these trade-offs is an important direction for future work.

\subsubsection{Prospects for Enhanced Skeleton Abstraction}
Our code-skeleton approach is straightforward: any bitwise difference yields a new key in the VectorDB. Alternative abstraction methods could produce more concise, expressive skeletons. Because this process is modular, experimenting with alternative abstraction techniques is simple and remains a promising direction for future research.

\subsubsection{Multiple Fix Generation, Differential Testing, and Reinforcement Learning}
\drfix{} currently deploys the first validated fix. While generating multiple candidates and using differential testing could improve fix accuracy, it also increases computational costs and time-to-solution. Moreover, \drfix{} does not use reinforcement learning to reward or penalize its outputs. Integrating such feedback could further improve the model’s capabilities over time.

\subsubsection{Open-source:}
We release code skeletons representing our data-race corpus along with selected data races and their fixes at:
\url{https://github.com/uber-research/drfix}.

\section{Related Work}
\label{sec:rel}
\paragraph{Data race detection:} Data race detection has been extensively studied, with a rich literature spanning diverse methodologies and applications~\citep{mellor1991fly, feng1997efficient,cheng1998detecting, ha2011efficient,ha2012fly,shi2024optimistic,kulkarni2021dynamic,kini2018data}.
Approaches typically include dynamic~\citep{o2003hybrid, flanagan2009fasttrack, smaragdakis2012sound, predictive_analysis_race, feng1997efficient, cheng1998detecting, savage1997eraser, michael_pldi_2018,yu2005racetrack, marino2009literace, raman2012scalable, sen2008race, serebryany2009threadsanitizer}, static \citep{mayur_pldi_2006, kahlon2007fast, racerd, pratikakis2011locksmith, chatarasi2017extended, engler2003racerx, voung2007relay}, or hybrid~\citep{atzeni2016archer, KasikciRaceMob} analysis. 
Our fixing approach can work with different race detection analyses in general, although {\it sound} race detection analyzes that provide stacktraces are preferred.

\noindent
\paragraph{Program Repair for Concurrency:}
There are many efforts on concurrency repair~\citep{lin2018pfix,liu2016understanding,liu2014grail,liu2012axis}.
To the best of our knowledge, the first works to tackle concurrency bugs automatically specialize in deadlock~\citep{yin_popl_2009,DBLP:conf/asplos/LiuZQCS21} and atomicity violation~\citep{jin2011automated} fixes. Furthermore, Jin et al. introduce CFix~\citep{jin2012automated} on top of the AFix work~\citep{jin2011automated} to support more concurrency bug types such as order violations. They introduce lock regions to repair the concurrency bugs and iteratively add more locks to resolve the introduced deadlock, if any.  Kelk et al. present ARC~\citep{kelk2013automatically}, a fully automated system designed to fix concurrency bugs, including data races and deadlocks in Java programs.
ARC operates in two phases: a repair phase, where genetic algorithms produce and test patches by modifying synchronization constructs, and an optimization phase, which eliminates unnecessary synchronization.
Liu et al.,~\citep{liu2021automatically} devise three strategies to fix channel-based deadlocks in Go programs.
Saioc et al.~\citep{saioc2025dynamic} extend the Go garbage collector to detect and remedy channel-based partial deadlocks.
Surendran et al.~\citep{surendran2014test} show a hybrid static-dynamic approach of fixing data races in task-parallel languages like Habanero Java and X10. The tool first detects data race dynamically with test input and then  inserts \texttt{finish} statements to eliminate these races.
Compared to the existing work, \drfix{} is not limited to a few well-known concurrency bug patterns; also, our fixing strategy goes far beyond just adding locks to avoid data races.

Transactional memory~\citep{HerlihyTM}, especially Hardware Transactional Memory (HTM)~\citep{YooTSX,RajwarTSX,DiceHTM}, has historically been considered effective at providing isolation against data races in concurrent programs. GOCC~\citep{ZhizhouGOCC} integrates HTM into Go applications. However, hardware support for HTM has diminished over time, primarily due to security vulnerabilities and associated mitigation efforts.
Recently, Saioc et al.~\citep{}

\noindent
\paragraph{Traditional Approach of Bug Fixing:}
Multiple research work have been proposed to automatically repair the program\citep{le2019automated,long2016automatic,nguyen2013semfix,weimer2009automatically}. 
For example, GenProg~\citep{le2011genprog} by Le Goues et al. employs genetic programming to generate program variants aimed at correcting software defects. 
In contrast, \drfix{} uses LLMs to explore the solution space while using RAG to efficiently narrow it down. Both GenProg and \drfix{} rely on the successful execution of existing test cases and do not require formal specifications or program annotations.
Additionally, Logozzo and Ball~\citep{logozzo2012modular} introduce a system that employs modular verification to automatically suggest program repairs during the design phase.
\noindent
\paragraph{GenAI-based Bug Fixing:}
Recently, there is  growing interest in using GenAI to fix bugs~\citep{wei2023copiloting,xia2023automated} with the advancement and popularity of the large language models (LLMs). Bouzenia et al. introduce RepairAgent~\citep{bouzenia2024repairagent}, an autonomous tool that uses LLM for program repair. Unlike traditional repair approaches with fixed prompts or static workflows, RepairAgent dynamically interacts with the code, autonomously chosses actions based on feedback from previous repairs. Jin et al.~\citep{jin2023inferfix} explored the idea of combining LLM with the Infer static analyzer to address critical bugs in Java and C\# including Null Pointer Dereference and Resource Leak. InferFix uses a retrieval-augmented prompting approach, where similar past fixes are retrieved to guide the model, which is then used to generate contextually relevant patches.
Xia and Zhang introduce ChatRepair~\citep{xia2024automated}, a conversational approach to Automated Program Repair (APR) using ChatGPT.
ChatRepair leverages dialogue-based interactions to iteratively generate and refine patches for buggy code.
Jin et al.~\citep{jinearly} apply data race fixes using GenAI on a small scale with a narrow use case.

None of these efforts have addressed concurrency bug fixing at an industry scale, where the effectiveness of GenAI approaches remains largely unexplored. 
Our work fills this gap. 
\drfix{} is the first to demonstrate strong effectiveness on complex, industry-scale Go codebases, where subtle and pervasive data races are common.

\section{Conclusions}
\label{sec:concl}
\drfix{} is a practical solution for addressing data races in large codebases, combining program analyses with large language models to automate fixes effectively.
\drfix{} leverages retrieval-augmented generation (RAG) and code skeletonization to retrieve structurally similar examples from a knowledge base, guiding the LLM to generate effective code patches.
Deployed at \company{}, \drfix{} demonstrated practical utility by addressing \numdrfixfixed{} data races over a \drfixperiod{} period, achieving an \fixeddrrate{} acceptance rate for developer-approved patches. 
This highlights the success of the tool in providing reliable and idiomatic fixes that align with developers' expectations while significantly reducing the average time required to address data races.
Although developed specifically for Go, the system’s principles were tested with other languages, suggesting potential for broader application. 
Overall, \drfix{} shows promise as a scalable tool for data race resolution in industry, contributing to ongoing efforts to improve code reliability in high-concurrency environments.

Our future work involves bridging the gap to reach near-perfect race fixes, which includes expanding to multiple files, external libraries, better bug categorization, and more concurrency patterns.
We also want to expand \drfix{} to fix bugs reported by other tools, including static analyses, and also to a wider set of programming languages.
It is also valuable to expand our approach to other kinds of concurrency bugs, such as deadlocks and atomicity violations.
Yet another stream of work is to employ LLMs to detect concurrency bugs and fix them in the existing codebase or during development. 

\bibliographystyle{ACM-Reference-Format}
\bibliography{references}
\clearpage
\appendix
\section{Loop Variable Capture Example}
\label{sec:loopvarcapture}

Listing~\ref{casestudy:exampleCaptureByReference} illustrates a case where the loop variable \texttt{num} is a free variable within the lambda body and is captured by reference on line~\ref{line:looprangecapture}. However, \texttt{num} is reassigned on each iteration of the loop at line~\ref{line:looprangeupdate}, referencing the value for the next iteration. This leads to a data race: while one iteration reads from \texttt{num}, another may simultaneously update it. Due to the prevalence of this issue, Go 1.22 revised the semantics of the \texttt{for}-loop to give loop variables per-iteration scope, rather than per-loop scope~\citep{go1.22:online}.
 
\begin{figure}
        \begin{lstlisting}[language=CustomGo, label={casestudy:exampleCaptureByReference}, caption={Data race due to capture-by-reference of loop variable and its fix by privatizing the value.}, captionpos=b]
nums := []int{0, 1, 2, 3, 4}
var wg sync.WaitGroup

@\label{line:looprangeupdate}@for _, num := range nums {
+++   num := num +++ // Create a private copy.
    wg.Add(1)
    go func() {
        defer wg.Done()
@\label{line:looprangecapture}@        fmt.Println(num)
    }()

}

wg.Wait()
        \end{lstlisting}
\end{figure}

\section{Concurrent Use of Thread-Unsafe \texttt{math/rand} APIs example}
\label{sec:miscrand}

Listing~\ref{casestudy:example8} illustrates a data race caused by concurrent access to a shared random number source in the Go standard library.
The handler function \mgo{responseRandom} generates a response body consisting of random bytes. Originally, it used a global \mgo{rand.Source} instance stored in the variable \mgo{\_responseRandomSource}. However, the Go \mgo{math/rand} package's \mgo{Source} and \mgo{Rand} types are not safe for concurrent use. When multiple HTTP requests are handled simultaneously, they may access this shared source concurrently, leading to a data race.

The fix is to eliminate the shared global source and instead initialize a new \mgo{rand.Source} for each request. This ensures that each \mgo{rand.Rand} instance is used by only one goroutine.
This example highlights the importance of understanding the concurrency guarantees (or lack thereof) in third-party and standard libraries. Even common packages can introduce subtle data races if used incorrectly in concurrent settings.
\begin{figure}[!t]
  \begin{lstlisting}[language=CustomGo,label={casestudy:example8},caption={Concurrent calls to thread-unsafe \mgo{math/rand} APIs and the corresponding fix.},captionpos=b]
@\rminus@ @\textcolor{red}{ var \_responseRandomSource = rand.NewSource(1001)}@
func (s *HTTP) responseRandom(w http.ResponseWriter, r *http.Request, p httprouter.Params) {
    ... // Rest of the code
    s.recordRequest(r, "response-random").Inc(1)
    size := getResponseSizeFromReq(r)
    if contentType := r.URL.Query().Get("content-type"); contentType != "" {
        w.Header().Set("Content-Type", contentType)
    }
@\rminus@   @\textcolor{red}{ random := rand.New(\_responseRandomSource)}@
@\gplus@   @\textcolor{darkgreen}{ random := rand.New(rand.NewSource(1001))}@
    io.CopyN(w, random, int64(size.Bytes()))
    ... // Rest of the code
}
  \end{lstlisting}
\end{figure}

\section{Pseudo Code Showing the Overall \drfix{} Algorithm}
\label{sec:pseudocode}
Listing~\ref{lst:fixgeneration} shows the overall approach of \drfix{} in fixing data races using LLMs. It employs race fixing strategy at various potential fixing locations at different granularities. 
It retrieves samples from a vector data base and also uses a feedback to the LLM showing any previous errors.
\lstset { %
    language=Python,
    basicstyle=\scriptsize\ttfamily,
    keywordstyle=\normalfont,
    morekeywords={},
    deletekeywords={int, return, if, else, for, while, map}
}
\begin{figure}[!t]
  \begin{lstlisting}[language=CustomPyton,label={lst:fixgeneration},caption={Algorithm for getting a data race fix from data race info.},captionpos=b]
def GetAFix(raceInfo, numRetries):
  for locationType in [TEST, LEAF, LCA]:
    for contentType in [FUNC, FILE]:
      for ex in GetExampleFromVectorDB(raceInfo, locationType, contentType, numExamples):
        lastFailure = ''
        for _ range(numRetries):
          prompt = GetPrompt(raceInfo, ex, locationType, contentType, lastFailure)
          patch = GetPatch(prompt)
          message, ok = ValidateFix(patch, raceInfo.bugHash)
          if ok:
            return patch # Produced a validated patch.
          lastFailure = lastFailure + message
  return None # Failed to produce patch.
\end{lstlisting}
\end{figure}

\section{Non-Trivial Data Race Fixes Enabled by RAG}
\label{sec:rag-example-fix}

In this section, we present a set of data race fixes that required deeper semantic understanding and context due to their non-trivial nature. These fixes were often complicated by the use of multiple synchronization constructs or subtle shared-state patterns. We show how \drfix{} leveraged Retrieval-Augmented Generation (RAG) to select semantically similar examples from the codebase, which in turn guided the LLM toward correct and robust fixes.

\subsection{RAG-Guided Fix via Parameter Passing}
\label{sec:goroutine-fix}

Listing~\ref{lst:data-race-fix} shows a fix where concurrent goroutines were inadvertently capturing the \texttt{data} object by reference. In the original version, one goroutine modified \texttt{data} via \texttt{SaveRate} while another read it via \texttt{SendRating}. The solution passes \texttt{data} as a parameter to each goroutine, ensuring that each operates on its own copy rather than sharing the same reference. Additionally, the \texttt{uploadImage} function signature was updated to use a pointer parameter to improve efficiency for read-only operations.
The above fix was guided by a similar pattern extracted via RAG (see Listing~\ref{lst:rag-example}). In that example, the goroutine originally captured both \texttt{ctx} and \texttt{request} by reference. The solution passed these as parameters to prevent unintended sharing, reinforcing the parameter passing strategy in such cases.

\begin{figure}[!t]
  \begin{lstlisting}[language=CustomGo,label={lst:data-race-fix},caption={Data race fix: passing \texttt{data} as parameters to goroutines to avoid capture-by-reference.},captionpos=b]
func (c *Controller) processRating(data Data, app App) {
    // ... existing code ...
    
@\rminus@   @\textcolor{red}{go func() \{}@
@\gplus@   @\textcolor{darkgreen}{go func(data Data) \{}@
      if issue := createIssue(data); issue.Valid {
          data.Status = "processed"
          _ = c.db.SaveRate(ctx, &data)
      }
@\rminus@   @\textcolor{red}{\}()}@
@\gplus@   @\textcolor{darkgreen}{\}(data)}@

@\rminus@ @\textcolor{red}{go func() \{}@
@\gplus@ @\textcolor{darkgreen}{go func(data Data) \{}@
      _ = c.notifier.SendRating(ctx, &data)
@\rminus@ @\textcolor{red}{\}()}@
@\gplus@ @\textcolor{darkgreen}{\}(data)}@
}
@\rminus@ @\textcolor{red}{func (c *Controller) uploadImage(img string, data Data, app App) \{}@
@\gplus@ @\textcolor{darkgreen}{func (c *Controller) uploadImage(img string, data *Data, app App) \{}@
    // ... existing code ...
}
 \end{lstlisting}
\end{figure}

\begin{figure}[!t]
  \begin{lstlisting}[language=CustomGo,label={lst:rag-example},caption={RAG example showing parameter passing in goroutines.},captionpos=b]
func processState(ctx context.Context, request *Request) {
 // ... existing code ...
    
@\rminus@   @\textcolor{red}{go func() \{}@
@\gplus@   @\textcolor{darkgreen}{go func(ctx context.Context, req *Request) \{}@
      result := processor.Execute(ctx, request)
      resultChan <- result
@\rminus@   @\textcolor{red}{\}()}@
@\gplus@   @\textcolor{darkgreen}{\}(ctx, request)}@
    // ... existing code ...
}
  \end{lstlisting}
\end{figure}

\subsection{RAG-Guided Fix for WaitGroup Synchronization}
\label{sec:wg-fix}

Listing~\ref{lst:fix-wg} demonstrates a fix for improper \texttt{WaitGroup} usage in a background worker handling state changes. 
The error involved calling \texttt{wg.Add(1)} inside the goroutine after performing channel operations and lock acquisitions. 
This sequencing could lead to a  race where the worker exits before the counter is incremented. 
The fix moves the \texttt{Add} call before launching the goroutine and utilizes \texttt{defer} to handle the counter's decrement. 
All coordination between channels, locks, and resource management remains intact.
A similar pattern was noted from RAG-based examples (see Listing~\ref{lst:rag-wg}). 
In that instance, the command executor coordinated multiple I/O copying goroutines correctly by sequencing the \texttt{WaitGroup} operations relative to channel and goroutine launches.

\begin{figure}[!t]
\begin{lstlisting}[language=CustomGo,label={lst:fix-wg},caption={WaitGroup synchronization fix: moving \texttt{wg.Add} before goroutine launch to prevent race.},captionpos=b]
func (w *Worker) processStateChanges() {
+++ w.wg.Add(1)+++
  go func() {
+++   defer w.wg.Done()+++
    for {
        select {
        case <-w.signalCh:
        case <-w.ctx.Done():
            return
        }

        w.mu.Lock()
        newState := w.state
        oldRes := w.currentResource()
        w.mu.Unlock()

        if cap(oldRes) == newState {
            continue
        }

        // State has changed, update resources
---       w.wg.Add(1)---
---       defer w.wg.Done()---
        newRes := allocateResource(newState)
        toTransfer := newState
        diff := newState - cap(oldRes)
        if diff > 0 {
            addResources(newRes, diff)
            toTransfer -= diff
        }
        w.setResource(newRes)
        transferResources(w.ctx, oldRes, toTransfer, func(r *resource) { 
                newRes.add(r) 
        })
        if diff < 0 {
            cleanupResources(w.ctx, oldRes, -diff)
        }
        releaseResource(oldRes)
    }
  }()
}
\end{lstlisting}
\end{figure}

\begin{figure}[!t]
\begin{lstlisting}[language=CustomGo,label={lst:rag-wg},caption={RAG example showing correct WaitGroup usage with multiple goroutines.},captionpos=b]
  func (c *CommandExecutor) ExecExistingCommand(cmdTimeout time.Duration, withTrace bool, command *exec.Cmd) (string, string, error) {
    timeout := time.After(cmdTimeout)
+++   var wg sync.WaitGroup+++
    var stdoutbytes, stderrbytes bytes.Buffer
    if withTrace {
        stdoutPipe, err := command.StdoutPipe()
        if err != nil {
            return "", "", err
        }
        stderrPipe, err := command.StderrPipe()
        if err != nil {
            return "", "", err
        }
        var stdoutWriter, stderrWriter io.Writer
        stdoutWriter = io.MultiWriter(os.Stdout, &stdoutbytes)
        stderrWriter = io.MultiWriter(os.Stderr, &stderrbytes)
+++       wg.Add(2)+++    
        go func() {
+++           defer wg.Done()+++
            io.Copy(stdoutWriter, stdoutPipe)
        }()
        go func() {
+++           defer wg.Done()+++
            io.Copy(stderrWriter, stderrPipe)
      }()
    } else {
        command.Stdout = &stdoutbytes
        command.Stderr = &stderrbytes
    }

    if err := command.Start(); err != nil {
        return "", "", errors.Wrap(err, "failed to start command")
    }

    done := make(chan error)
    go func() { done <- command.Wait() }()

    select {
        case <-timeout:
            command.Process.Kill()
            return "", "", errors.New("command timed out")
        case err := <-done:
+++           wg.Wait()+++
            return stdoutbytes.String(), stderrbytes.String(), err
    }
}
\end{lstlisting}
\end{figure}

\subsection{RAG-Guided Fix for Concurrent Map Access}
\label{sec:syncmap-fix}
Listing~\ref{lst:fix-map} addresses data races encountered during concurrent map accesses. 
The original implementation used a standard map to track locks across shards, which could lead to unsafe concurrent modifications. 
The fix replaces the map with \texttt{sync.Map} and revises the operations accordingly. 
This ensures that all operations are thread-safe.
This approach was similarly validated by a RAG-based example (see Listing~\ref{lst:rag-map}), where a test helper replaced the map with \texttt{sync.Map} to safely track retry counts in concurrent test runs.
The shown change and the applied change both change the variable type and modify the APIs applied to the type, which are scattered throughout the file.

\begin{figure}[!t]
\begin{lstlisting}[language=CustomGo,label={lst:fix-map},caption={Fixing concurrent map access by replacing map with sync.Map.},captionpos=b]
type Manager struct {
---   items     map[Key]Item---
+++   items     sync.Map+++
    // ... existing code ...
}

func (m *Manager) Initialize() *Manager {
---   m.items = make(map[Key]Item)---
+++   m.items = sync.Map{+++}
    // ... existing code ...
}

func (m *Manager) cleanup() {
    // Remove unnecessary items
---   for key := range m.items {---
---       if !m.isValid(key) {---
---           delete(m.items, key)---
---       }---
---   }---
+++   m.items.Range(func(key, value interface{}) bool {+++
+++       if !m.isValid(key.(Key)) {+++
+++           m.items.Delete(key)+++
+++       }+++
+++       return true+++
+++   })+++

    // Process new items
    for key := range newItems {
---       if _, exists := m.items[key]; !exists {---
---           m.items[key] = Item{---
---               mutex: &sync.Mutex{},---
---               active: false,---
---           }---
---       }---
---       item := m.items[key]---
+++       val, _ := m.items.LoadOrStore(key, Item{+++
+++           mutex: &sync.Mutex{},+++
+++           active: false,+++
+++       })+++
+++       item := val.(Item)+++
        go m.process(key, item)
    }
}
\end{lstlisting}
\end{figure}

\begin{figure}[!t]
\begin{lstlisting}[language=CustomGo,label={lst:rag-map},caption={RAG example showing conversion from map to sync.Map for concurrent access.},captionpos=b]
---var retryCount = map[string]int{}---
+++var retryCount = sync.Map{}+++
func handleRequest(w http.ResponseWriter, r *http.Request) {
---   if count, exists := retryCount[r.URL.Path]; exists {---
---       retryCount[r.URL.Path] = count + 1---
---   } else {---
---       retryCount[r.URL.Path] = 1---
---   }---
+++   if val, ok := retryCount.Load(r.URL.Path); ok {+++
+++       retryCount.Store(r.URL.Path, val.(int) + 1)+++
+++   } else {+++
+++       retryCount.Store(r.URL.Path, 1)+++
+++   }+++

---   if retryCount[r.URL.Path] != maxRetries {---
+++       if count, _ := retryCount.Load(r.URL.Path); count != maxRetries {+++
            time.Sleep(retryDelay)
        }
    }
    ...
\end{lstlisting}
\end{figure}

\subsection{RAG-Guided Fix for Atomic Operations}
\label{sec:atomic-fix}
Listing~\ref{lst:fix-atomic} presents a fix for data races arising from concurrent integer access. 
The original code used a plain integer variable to track the batch length, which could lead to race conditions. 
By switching to atomic operations, the fix ensures thread-safe updates and reads while preserving the intricate synchronization logic between goroutines (including channel coordination, RWMutex locks, and WaitGroups).
This change was reinforced by a RAG example (see Listing~\ref{lst:rag-atomic}), where a test server applied atomic operations to safely update concurrent request counters.

\begin{figure}[!t]
\begin{lstlisting}[language=CustomGo,label={lst:fix-atomic},caption={Fixing concurrent integer access using atomic operations in batch processing.},captionpos=b]
func FetchAll(
    ctx context.Context,
    countPerRequest int,
    startNo int,
    endNo int,
    batchCount int,
) ([]entity.User, error) {
    users := make([]entity.User, 0)
    sendRequestChan := make(chan struct{})
    stopRequestChan := make(chan struct{})
    doneChan := make(chan struct{})
    errChan := make(chan error)
    responseChan := make(chan []entity.User)
---   minBatchResponseLength := countPerRequest---
+++   minBatchResponseLength := int32(countPerRequest)+++
    wg := new(sync.WaitGroup)
    waitRWMutex := sync.RWMutex{}

    var err error
    // trigger parallel batches of request, stops in case of error or empty results returned
    go func() {
        for {
            select {
                case err = <-errChan:
                ...
                case <-stopRequestChan:
                    close(sendRequestChan)
                    close(responseChan)
                    close(doneChan)
                    return
                case response := <-responseChan:
                    if startNo > endNo {
---                       minBatchResponseLength = 0---
+++                       atomic.StoreInt32(&minBatchResponseLength, 0)+++
                    }
                    users = append(users, response...)
                    case <-sendRequestChan:
                    for i := 0; i < batchCount; i++ {
                        waitRWMutex.Lock()
                        wg.Add(1)
                        waitRWMutex.Unlock()
                        go getInfoFromDB(
                        ctx, wg, responseChan, errChan, startNo, startNo+countPerRequest-1, endNo)
                        startNo += countPerRequest
                    } // for
            } // select
        } //for 
    }()

    // a checker to ensure whether next batch of requests should be sent or not
    go func() {
        for {
---           switch minBatchResponseLength {---
---               case 0:--- // close and return.
---                    close(stopRequestChan)---
---                    return---
---               default:---
---                   sendRequestChan <- struct{}{}---// send the next batch.
---           }---
+++           if atomic.LoadInt32(&minBatchResponseLength) == 0 {+++
+++                close(stopRequestChan)+++ // close and return.
+++                return+++
+++           }  +++
+++           sendRequestChan <- struct{}{}+++ // send the next batch.
            waitRWMutex.RLock()
            wg.Wait()
            waitRWMutex.RUnlock()
        }
    }()
    <-doneChan
    defer close(errChan)
    if err != nil {
        return users, err
    }
    return users, nil
}
\end{lstlisting}
\end{figure}

\begin{figure}[!t]
\begin{lstlisting}[language=CustomGo,label={lst:rag-atomic},caption={RAG example showing conversion to atomic operations for concurrent counter.},captionpos=b]
func newServer(t *testing.T, responses []string) (*url.URL, *httptest.Server) {
---   cnt := 0---
+++   var cnt int32+++
    s := svr.NewServer(http.HandlerFunc(func(w http.ResponseWriter, r *http.Request) {
        w.WriteHeader(http.StatusOK)
---       if cnt > 0 {---
+++           if atomic.LoadInt32(&cnt) > 0 {+++
                io.WriteString(w, responses[1])
            } else {
                io.WriteString(w, responses[0])
            }
---           cnt++---
+++           atomic.AddInt32(&cnt, 1)+++
        }))
        u, err := url.Parse(s.URL)
        return u, s
    }
    ... // rest of the code
}    
\end{lstlisting}
\end{figure}

\subsection{RAG-Guided Fix for Struct Copy}
\label{sec:struct-copy-fix}

Listing~\ref{lst:struct-copy-fix} deals with data races in struct access. 
The original code modified shared configuration structures directly, which could lead to race conditions. 
The fix creates a copy of the struct, applies the necessary modifications on the copy, and then passes that copy along. 
This method prevents concurrent modifications of the shared configuration while preserving functionality.
RAG example (see Listing~\ref{lst:struct-copy-rag}) presents a similar situation where a recommendation service creates a copy of the base request struct before applying modifications to avoid concurrent races.

\begin{figure}[!t]
\begin{lstlisting}[language=CustomGo,label={lst:struct-copy-fix},caption={Fixing concurrent struct access by copying before modification.},captionpos=b]
func defaultNewConsumerFn(
    config *service.Config,
    brokers []string,
) (Consumer, error) {
---   config.Consumer.Return.Errors = true---
---   config.Consumer.MaxProcessingTime = 60 * time.Second---
---   return service.NewConsumer(brokers, config)---
+++   newConfig := *config+++
+++   newConfig.Consumer.Return.Errors = true+++
+++   newConfig.Consumer.MaxProcessingTime = 60 * time.Second+++
+++   return service.NewConsumer(brokers, &newConfig)+++
}
\end{lstlisting}
\end{figure}

\begin{figure}[!t]
\begin{lstlisting}[language=CustomGo,label={lst:struct-copy-rag},caption={RAG example showing struct copy to prevent concurrent modification.},captionpos=b]
func (r *RecommendationsImpl) buildRecommendationRequest(
    ctx context.Context, 
    request *Request, 
    geofence *Geofence) *RecommendationRequest {
    
    baseRequest, _ := r.feedClient.MakeBaseGetFeedRequest(ctx)
    if baseRequest == nil {
        return nil
    }
---   request.CityID = &geofence.CityID---
---   request.CityName = &geofence.CityName---
---   request.TargetTimezone = &geofence.Timezone---
+++   copyRequest := *baseRequest+++
+++   copyRequest.CityID = &geofence.CityID+++
+++   copyRequest.CityName = &geofence.CityName+++
+++   copyRequest.TargetTimezone = &geofence.Timezone+++

---   request.TargetLocation = &Location{---
+++   copyRequest.TargetLocation = &Location{+++
        Latitude:  request.TargetLocationLatitude,
        Longitude: request.TargetLocationLongitude,
    }
    req := &RecommendationRequest{
---       FeedRequest:    request,---
+++       FeedRequest:    &copyRequest,+++
            ...
    }
    return req
}

\end{lstlisting}
\end{figure}

\subsection{RAG-Guided Fix for Struct Copy in Complex Lambda Return}
\label{sec:struct-copy-lambda-fix}

Listing~\ref{lst:struct-copy-lambda} demonstrates a fix for a data race in a scenario where a lambda function captured and later modified an internal struct field. 
Instead of modifying the field directly, the lambda now creates a copy before making updates. 
This design preserves the integrity of the original shared state while allowing safe modifications within the closure.
This strategy was informed by a similar case via RAG (see Listing~\ref{lst:struct-copy-promo}), where a controller made a copy of \texttt{fareInfo} before concurrently dispatching processing functions, ensuring that shared state was not modified concurrently.

\begin{figure}[!t]
\begin{lstlisting}[language=CustomGo,label={lst:struct-copy-lambda},caption={Fixing concurrent struct access in lambda execution by copying before modification.},captionpos=b]
func BuildUpdatedFulfillmentType(
    fetchAccountJob *accountjobs.FetchAccountJob,
    deliverabilityResult *datamodels.Response,
) func() (stores.StoreAdapter, bool) {
    return func() (wrapper stores.StoreAdapter, ok bool) {
        account := fetchAccountJob.Account

        if isValid := isValidStoreAcct(account); !isValid {
            return nil, false
        }

+++       internalAccountCopy := *account.InternalAccount+++

---       storeUUID := string(*account.InternalAccount.UUID) ---
+++       storeUUID := string(*internalAccountCopy.UUID) +++
        if deliverabilityResult != nil &&
        deliverabilityResult.DeliverabilityMap != nil {
            deliverability := deliverabilityResult.DeliverabilityMap[bazr.UUID(storeUUID)]
            if deliverability != nil {
                // Update the copy with the new fulfillment type
---               account.InternalAccount.FulfillmentType = deliverability.FulfillmentType---
+++               internalAccountCopy.FulfillmentType = deliverability.FulfillmentType+++
            }
        }
        return &stores.InternalAccountWrapper{
---           Base: *account,---
+++           Base: internalAccountCopy,+++
        }, true
    }
}
\end{lstlisting}
\end{figure}

\begin{figure}[!t]
\begin{lstlisting}[language=CustomGo,label={lst:struct-copy-promo},caption={RAG example showing struct copy to prevent races in captured variables.},captionpos=b]
func (c *controller) GetPromoTrackings(
    ctx context.Context,
    request *Request,
) (*Response, error) {
    var fareInfo *entity.FareInfo
    var err error

    var fareInfoFromGateway *entity.FareInfo
    var fareEstimateFromGatewayErr error
    fareInfoFromGateway, fareEstimateFromGatewayErr = mapper.MapFareEstimateToFareInfo(...)

    if request.FareSessionUUID != nil {
        if fareInfo, err = c.getFareInfo(...); err != nil {
            return nil, err
        }

+++       fareInfoCopy := *fareInfo+++
+++       errCopy := err+++

        safego.Go(func() { // do aysnc compare
---           migration.DoPromoTracking(ctx, request, fareInfo, fareInfoFromGateway, err, fareEstimateFromGatewayErr)---
+++           migration.DoPromoTracking(ctx, request, &fareInfoCopy, fareInfoFromGateway, errCopy, fareEstimateFromGatewayErr)+++
        })

        safego.Go(func() { // do aysnc compare
---           migration.DoPromoTracking(ctx, request, fareInfo, fareInfoFromGateway, err, fareEstimateFromGatewayErr)---
+++           migration.DoPromoTracking(ctx, request, &fareInfoCopy, fareInfoFromGateway, errCopy, fareEstimateFromGatewayErr)+++
        })
        // in reverse shadow, we will use other fare info as source of truth
        fareInfo = fareInfoFromOrderGateway // overwrites fareInfo

    } else {
        fareInfo = c.getFareInfoWithoutFareSession(ctx, request) 
    }
    // overwrites err
    err = c.getPromotionsAndContext(...)
    // rest of the code ...
}
\end{lstlisting}
\end{figure}

\subsection{RAG-Guided Fix for Struct Copy in Error Handling}
\label{sec:struct-copy-error-fix}

Listing~\ref{lst:struct-copy-error} shows a fix for a data race in error handling. The original \texttt{WithSourceError} method modified the \texttt{error} field of the receiver directly, risking race conditions when shared among goroutines. The fix creates a copy of the base error, sets the source error on the copy, and returns the new object. This change ensures that error objects remain immutable and thread-safe.
While this is sematically not the same as setting a field of the object, but the developers admitted this change since the objective was to get an error object.
Another RAG example (see Listing~\ref{lst:struct-copy-carrier}) reinforces this approach. In that case, the function creates a copy of a shared response object (including its nested \texttt{Carrier} field), performs the necessary modifications, and returns the safe copy.

\begin{figure}[!t]
\begin{lstlisting}[language=CustomGo,label={lst:struct-copy-error},caption={Fixing concurrent error handling by copying before modification.},captionpos=b]
func (e *GatewayErrorImpl) 
func (e *GatewayErrorImpl) WithSourceError(sourceErr error) SolarGateway {
---   e.error = sourceErr---
---   return e---
+++   newError := *e+++
+++   newError.error = sourceErr+++
+++   return &newError+++
}
\end{lstlisting}
\end{figure}

\begin{figure}[!t]
\begin{lstlisting}[language=CustomGo,label={lst:struct-copy-carrier},caption={RAG example showing copying of shared response object.},captionpos=b]
func (c *controller) fetchCarrierInfo(ctx context.Context, carrierUUID string) (*user.FetchCarrierRes, error) {
    res, err := c.ufo.FetchCarrier(ctx, carrierUUID)  
    ...

    if res.Carrier.AccountManager == nil || res.Carrier.AccountManager.Email == nil {
+++       resCopy := &user.FetchCarrierRes{}+++
+++       *resCopy = *res+++
+++       resCopy.Carrier = &shared.Carrier{}+++
+++       *resCopy.Carrier = *res.Carrier+++
---       res.Carrier.AccountManager = &shared.Operator{Email: ptr.String("")}---
+++       resCopy.Carrier.AccountManager = &shared.Operator{Email: ptr.String("")}+++
+++       return resCopy, nil+++
    }
    return res, nil
}
\end{lstlisting}
\end{figure}

\subsection{RAG-Guided Fix for Initialization Sequence}
\label{sec:scheduler-init-fix}

Listing~\ref{lst:struct-copy-scheduler} addresses a race during scheduler initialization. In the original design, the scheduler started in a goroutine while dependent goroutines were launched immediately after, risking premature access. The fix introduces a \texttt{readyChan} to signal when the scheduler is fully initialized before any dependent goroutines begin execution. Although a synchronous start of the scheduler was an alternative, the chosen channel-based approach preserves asynchrony while ensuring proper synchronization.
This solution was validated by a RAG example (see Listing~\ref{lst:struct-copy-provision}) that replaced brittle time-based waiting with channel-based synchronization, leading to more reliable and deterministic behavior.

\begin{figure}[!t]
\begin{lstlisting}[language=CustomGo,label={lst:struct-copy-scheduler},caption={Scheduler initialization fix using ready channel.},captionpos=b]
    ...
    // start the scheduler in another goroutine
+++   readyChan := make(chan struct{})+++
    go func() {
        scheduler.Start(context.Background())
+++       close(readyChan)+++
          ...
    }()
+++   <-readyChan+++

    wg.Add(len(tt.calls))
    for i := 0; i < len(tt.calls); i++ {
        go func(idx int) {
            defer wg.Done()
            err := scheduler.QueueAll(tt.calls[idx])
            assert.Equal(t, tt.expectErr[idx], err)
        }(i)
    }
    wg.Wait()
    scheduler.Stop(context.Background())
    ...
\end{lstlisting}
\end{figure}

\begin{figure}[!t]
\begin{lstlisting}[language=CustomGo,label={lst:struct-copy-provision},caption={RAG example showing channel-based test synchronization.},captionpos=b]
func TestProvisionErrorUndo(t *testing.T) {
    doCalled := false
---   undoCalled := false---
+++   undoCalled := make(chan bool)+++

    provisionStep := &provisionMemberStep{
        metricsScope: metrics.None,
        steps: []provisionStep{
            newDummyProvisionStep(
            "good",
            func() error {
                doCalled = true
                return nil
            },
---           func() { undoCalled = true }),---
+++           func() { undoCalled <- true }),+++
            newDummyProvisionStep(
            "bad",
            func() error {
                return errors.New("oops")
            },
            func() { assert.Fail(t, "don't call") }),
            newDummyProvisionStep(
            "dont call",
            func() error {
                assert.Fail(t, "don't call")
                return nil
            },
            func() { assert.Fail(t, "don't call") }),
        },
    }
    time.Sleep(time.Millisecond * 100)
    err := provisionStep.do(buildProvisionContext())

    assert.Error(t, err)
    assert.True(t, doCalled)
---   assert.True(t, undoCalled)---
+++   assert.True(t, <-undoCalled)+++
}
\end{lstlisting}
\end{figure}

\subsection{RAG-Guided Fix for Partial Locking in State Management}
\label{sec:db-fix}

In Listing~\ref{lst:struct-copy-db}, a subtle data race existed due to partial lock protection over shared state. 
The problem was compounded by multiple synchronization mechanisms—such as mutexes, waitgroups, and channels—that interact in non-trivial ways. 
Its complexity lies in the fact that, prior to the fix, the state was being accessed inconsistently from various parts of the code—some protected by locks and others left unguarded.
The following fix addresses these issues by introducing comprehensive mutex protection using a \texttt{sync.RWMutex}. 
This approach guarantees that every access to the shared map \texttt{lastProcessedWindow} is appropriately synchronized: write operations acquire the exclusive lock, and read operations leverage a read lock, reducing the window for race conditions and ensuring data integrity.
The RAG-guided example presented in Listing~\ref{lst:struct-copy-hit} and~\ref{lst:struct-copy-hit2} show addressing a data race condition in the \texttt{MonJob} type caused by unsynchronized access to shared state. The race was complicated by the concurrent nature of the system: channels were used to signal when the processing should stop and workgroups (\texttt{sync.WaitGroup}) managed the concurrent execution of the \texttt{Ping} method. These mechanisms led to scenarios where multiple goroutines could simultaneously access and modify the \texttt{jobExist} and \texttt{outputReceived} flags. To resolve this issue, a \texttt{sync.Mutex} was added to the \texttt{MonJob} structure and used in several functions to protect access to these shared variables. In the \texttt{Prepare} function, the mutex is locked while the flags are initialized. Within the \texttt{Ping} function, the mutex is used to safely capture the state of \texttt{jobExist} and to protect updates to both \texttt{jobExist} and \texttt{outputReceived} when checking timeouts and conditions. The same principle applies in \texttt{tryStartingJob} and \texttt{handleMessage}.
This example guided the LLM to look for lock protecting the shared variables at different code locations in a consistent manner.

\begin{figure}[!t]
\begin{lstlisting}[language=CustomGo,label={lst:struct-copy-db},caption={Fixing concurrent map access in database tiering job using RWMutex.},captionpos=b]
type dbTableDataTieringJob struct {
    tableConcurrency    int
    lastProcessedWindow map[int]string
---   mutex               sync.Mutex---
+++   mutex               sync.RWMutex+++
}

func (j *dbTableDataTieringJob) Run(ctx context.Context) error {
    var wg sync.WaitGroup
    concurrency := make(chan struct{}, j.tableConcurrency)

    var firstErr error
    var mu sync.Mutex // Protects firstErr only

+++   j.mutex.Lock()+++
    for k := range j.lastProcessedWindow {
        delete(j.lastProcessedWindow, k)
    }
+++   j.mutex.Unlock()+++

    tables := []int{101, 202}
    for _, tbl := range tables {
        wg.Add(1)
        go func(tableID int) {
            defer wg.Done()

            concurrency <- struct{}{}
            defer func() { <-concurrency }()

---           windowData := j.lastProcessedWindow[tableID]--- 
+++           j.mutex.RLock()+++
+++           windowData := j.lastProcessedWindow[tableID]+++
+++           j.mutex.RUnlock()+++

            err := doWork(windowData)
            if err != nil {
                mu.Lock()
                if firstErr == nil {
                    firstErr = err
                }
                mu.Unlock()
                return
            }

            j.mutex.Lock()
            j.lastProcessedWindow[tableID] = "some new data"
            j.mutex.Unlock()

        }(tbl)
    }

    wg.Wait()
    return firstErr
}
\end{lstlisting}
\end{figure}

\begin{figure}[!t]
\begin{lstlisting}[language=CustomGo,label={lst:struct-copy-hit},caption={Fix for partial locking in state management.},captionpos=b]

// MonJob defines the job that pings the system monitored by plugin.
type MonJob struct {
    ...
    jobExist       bool 
    outputReceived bool
+++   mutex sync.Mutex+++
}

func (hj *MonJob) Prepare(cfg *config.Config) error {
    ...
+++   hj.mutex.Lock()+++
    hj.jobExist = false
    hj.outputReceived = false
+++   hj.mutex.Unlock()+++
    ...
}
// Run is an infinite loop. The only requirement is that it stops when
// the `stop` channel is closed. The channel can be closed in multiple cases:
func (hj *MonJob) Run(stop chan struct{}) {
    wg := new(sync.WaitGroup)
    MonJobLoop:
    for {
        select {
            default:
            hj.bucket.Wait(1)

            wg.Add(1)
            go func() {
                defer wg.Done()
                hj.Ping()
            }()
            case <-stop:
            break MonJobLoop
        }
    }
    wg.Wait()
}
// tryStartingJob tries to start an AthenaX job.
// The start result will be recorded as a metric suggesting the platform availability
func (hj *MonJob) tryStartingJob(ctx context.Context) (err error) {
    if cond {
+++       hj.mutex.Lock()+++
        hj.jobExist = true
+++       hj.mutex.Unlock()+++
        return nil
    }
    ...
}

// handleMessage is the callback function of handler consumer.
func (hj *MonJob) handleMessage(ctx context.Context, message consumer.Message) error {
+++   hj.mutex.Lock()+++
+++   jobExist := hj.jobExist+++
+++   outputReceived := hj.outputReceived+++
+++   hj.mutex.Unlock()+++
---   if hj.jobExist && !hj.outputReceived {--- 
+++   if jobExist && !outputReceived {+++
+++       hj.mutex.Lock()+++
        hj.outputReceived = true
+++       hj.mutex.Unlock()+++
        ...
    }
    return nil
}

\end{lstlisting}
\end{figure}

\begin{figure}[!t]
\begin{lstlisting}[language=CustomGo,label={lst:struct-copy-hit2},caption={Fix for partial locking in state management continued.},captionpos=b]
// Ping is where the actual work happens.
func (hj *MonJob) Ping() {
    ctx, cancel := context.WithTimeout(context.Background(), 3*time.Minute)
    defer cancel()

+++   hj.mutex.Lock()+++
+++   jobExist := hj.jobExist+++
+++   hj.mutex.Unlock()+++
---   if hj.jobExist {--- 
+++   if jobExist {+++
        // If timeout reached before receiving any  output, we assume job starting failed
        if time.Since(hj.startTime) > hj.timeOut {
            ...
            if cond1 {
+++               hj.mutex.Lock()+++
+++               outputReceived := hj.outputReceived+++
+++               hj.mutex.Unlock()+++
---               if !hj.outputReceived {--- 
+++               if outputReceived {+++
                    ...
                }   
                ...
+++               hj.mutex.Lock()+++
                hj.jobExist = false
                hj.outputReceived = false
+++               hj.mutex.Unlock()+++
            } else if cond2 {
+++               hj.mutex.Lock()+++
+++               outputReceived := hj.outputReceived+++
+++               hj.mutex.Unlock()+++
---               if !hj.outputReceived {--- 
+++               if !outputReceived {+++
                    ...
                }
                ...
+++               hj.mutex.Lock()+++
                hj.jobExist = false
                hj.outputReceived = false
+++               hj.mutex.Unlock()+++
            }

        }
    } else if cond3 {
        hj.Exec(func() error {
            err := hj.tryStartingJob(ctx)
+++           hj.mutex.Lock()+++
            hj.jobExist = true
+++           hj.mutex.Unlock()+++
            ...
            return err
        })
    }
}
\end{lstlisting}
\end{figure}

\clearpage
\section{Actual Prompt}
\label{sec:prompt}
\begin{mdframed}[backgroundcolor=brown!20,linecolor=red!40!black,linewidth=3pt]

\{

    "role": "system",
  
    "content": "You are an expert in parallel computing and helping user fix data race in the golang programs. The user will provide you code delimited by the <code> </code> XML tag; you will try to fix the race. Your response should only contain the fixed code. Pay strong attention to the following instructions:
    
    (1) Do not skip any code by saying `the rest of the code stays the same` or `... rest of test cases ...` or `// Test cases...`.
    
    (2) Your response should be the entire revised code top to bottom, verbatim. Do not say any other thing.
    
    (3) Do not wrap the code with ```go``` or ```<code>```.
    
    (4) Absolutely, do not update or remove existing comments in the code.

\}, 

\{

  "role": "user",
  
  "content": "Refactor the code within <code> </code> XML tags to fix data race in golang function. You will be given 1 example that fix data race in golang function.
  
    Example 0 (Code with data race):

        ```code with data race```

    Example 0 (Code after data race):

        ```code after data race```

    The data race happens due to a memory conflict on a shared variable Read on line ```line1``` with the same shared variable write on line ```line2``` in <code> code to fix </code>.
    
\}

\end{mdframed}

\end{document}